\documentclass[
superscriptaddress,
nofootinbib,
twocolumn,
 amsmath,amssymb,
 aps,
pra,
notitlepage,
longbibliography
]{revtex4-1}

\usepackage{dcolumn}
\usepackage{bm}
\usepackage{epsfig}
\usepackage{graphicx}
\usepackage{latexsym}
\usepackage{amsfonts}
\usepackage{setspace}
\usepackage{graphicx}
\usepackage{amsmath}
\usepackage{verbatim}
\usepackage{color}
\usepackage{SIunits}
\usepackage{hyperref}
\usepackage{nccmath}

\newcommand{\be}{\begin{equation}}
\newcommand{\ee}{\end{equation}}
\newcommand{\ba}{\begin{array}}
\newcommand{\ea}{\end{array}}
\newcommand{\bqa}{\begin{eqnarray}}
\newcommand{\eqa}{\end{eqnarray}}
\newcommand{\ev}[1]{\ensuremath{\left\langle #1 \right\rangle}}

\renewcommand{\Re}{\text{Re}}

\newcommand{\ncavO}{n_{\text{c}}}
\newcommand{\gzeroO}{g_{\text{0}}}

\newcommand{\kappai}{\kappa_{\text{i}}}
\newcommand{\kappae}{\kappa_{\text{e}}}

\newcommand{\omegacO}{\omega_{\text{c}}}
\newcommand{\omegamO}{\omega_{\text{m}}}
\newcommand{\omegapO}{\omega_{\text{p}}}

\newcommand{\ahat}{\hat{a}}
\newcommand{\adag}{\hat{a}^{\dagger}}
\newcommand{\akhat}{\hat{a}_{k}}
\newcommand{\akdag}{\hat{a}^{\dagger}_{k}}
\newcommand{\aLhat}{\hat{a}_{\text{L}}}
\newcommand{\aLdag}{\hat{a}^{\dagger}_{\text{L}}}
\newcommand{\aRhat}{\hat{a}_{\text{R}}}
\newcommand{\aRdag}{\hat{a}^{\dagger}_{\text{R}}}
\newcommand{\bhat}{\hat{b}}
\newcommand{\bdag}{\hat{b}^{\dagger}}
\newcommand{\bWhat}{\hat{b}_{\text{W}}}
\newcommand{\bWdag}{\hat{b}^{\dagger}_{\text{W}}}
\newcommand{\bW}{b_{\text{W}}}
\newcommand{\bWprime}{b_{\text{W}^{\prime}}}
\newcommand{\bL}{b_{\text{L}}}
\newcommand{\bR}{b_{\text{R}}}
\newcommand{\tL}{t_{\text{L}}}
\newcommand{\tR}{t_{\text{R}}}
\newcommand{\dhat}{\hat{d}}
\newcommand{\ddagg}{\hat{d}^{\dagger}}

\newcommand{\OLR}{O_{\text{L(R)}}}
\newcommand{\OL}{O_{\text{L}}}
\newcommand{\OR}{O_{\text{R}}}
\newcommand{\lambdaLR}{\lambda_{\text{L(R)}}}

\newcommand{\kappaLR}{\kappa_{\text{L(R)}}}
\newcommand{\kappaL}{\kappa_{\text{L}}}
\newcommand{\kappaR}{\kappa_{\text{R}}}
\newcommand{\kappaiLR}{\kappa_{\text{iL(R)}}}

\newcommand{\kappaeLR}{\kappa_{\text{eL(R)}}}
\newcommand{\kappaeL}{\kappa_{\text{eL}}}
\newcommand{\kappaeR}{\kappa_{\text{eR}}}
\newcommand{\omegacL}{\omega_{\text{cL}}}
\newcommand{\omegacR}{\omega_{\text{cR}}}
\newcommand{\omegas}{\omega_{\text{s}}}
\newcommand{\omegap}{\omega_{\text{p}}}
\newcommand{\omegac}{\omega_{\text{c}}}
\newcommand{\omegack}{\omega_{\text{c},k}}
\newcommand{\omegamod}{\omega_{\text{mod}}}
\newcommand{\Gammapm}{\Gamma_{\pm}}
\newcommand{\Apm}{A_{\pm}}

\newcommand{\Gk}{G_{{k}}}

\newcommand{\GL}{G_{\text{L}}}
\newcommand{\GR}{G_{\text{R}}}
\newcommand{\PhiB}{\Phi_{\text{B}}}

\newcommand{\phiL}{\phi_{\text{L}}}
\newcommand{\phiR}{\phi_{\text{R}}}
\newcommand{\ncL}{n_{\text{cL}}}
\newcommand{\ncR}{n_{\text{cR}}}
\newcommand{\TLR}{T_{\text{R}\rightarrow\text{L}(\text{L}\rightarrow\text{R})}}
\newcommand{\TL}{T_{\text{R}\rightarrow\text{L}}}
\newcommand{\TR}{T_{\text{L}\rightarrow\text{R}}}

\newcommand{\gzeroL}{g_{0,\text{L}}}
\newcommand{\gzeroR}{g_{0,\text{R}}}
\newcommand{\gzeroLR}{g_{0,\text{L(R)}}}
\newcommand{\gzeroWL}{g_{0,\text{WL}}}
\newcommand{\gzeroWR}{g_{0,\text{WR}}}
\newcommand{\gzeroWk}{g_{0,\text{W}k}}

\newcommand{\GWL}{G_{\text{WL}}}
\newcommand{\GWR}{G_{\text{WR}}}
\newcommand{\Ppbar}{\bar{P}_{\text{p}}}
\newcommand{\PpL}{P_{\text{pL}}}
\newcommand{\PpR}{P_{\text{pR}}}

\newcommand{\Mw}{M_{\text{W}}}
\newcommand{\Mpm}{M_{\pm}}
\newcommand{\MLR}{M_{\text{L(R)}}}
\newcommand{\ML}{M_{\text{L}}}
\newcommand{\MR}{M_{\text{R}}}
\newcommand{\gammaiLR}{\gamma_{\text{iL(R)}}}
\newcommand{\gammaik}{\gamma_{{ik}}}
\newcommand{\gammaiL}{\gamma_{\text{iL}}}
\newcommand{\gammaiR}{\gamma_{\text{iR}}}
\newcommand{\omegamL}{\omega_{\text{mL}}}
\newcommand{\omegamR}{\omega_{\text{mR}}}
\newcommand{\omegamW}{\omega_{M_\text{W}}}
\newcommand{\omegamWprime}{\omega_{M_\text{W}^{\prime}}}

\newcommand{\gammaiW}{\gamma_{\text{iW}}}
\newcommand{\gammai}{\gamma_{\text{i}}}
\newcommand{\omegam}{\omega_{\text{m}}}

\begin{document}

\title{Generalized nonreciprocity in an optomechanical circuit via synthetic magnetism and reservoir engineering}

\author{Kejie Fang} 
\author{Jie Luo}
\affiliation{Kavli Nanoscience Institute, California Institute of Technology, Pasadena, California 91125, USA}
\affiliation{Institute for Quantum Information and Matter and Thomas J. Watson, Sr., Laboratory of Applied Physics, California Institute of Technology, Pasadena, California 91125, USA}
\author{Anja Metelmann}
\affiliation{Department of Physics, McGill University, 3600 rue University, Montr\'{e}al, Quebec H3A 2T8, Canada}
\affiliation{Department of Electrical Engineering, Princeton University, Princeton, New Jersey 08544, USA}
\author{Matthew H. Matheny}
\affiliation{Kavli Nanoscience Institute, California Institute of Technology, Pasadena, California 91125, USA}
\affiliation{Department of Physics, California Institute of Technology, Pasadena, California 91125, USA}
\author{Florian Marquardt}
\affiliation{Max Planck Institute for the Science of Light, G\"unther-Scharowsky-Stra\ss e 1/Bau 24, 91058 Erlangen, Germany}
\affiliation{Institute for Theoretical Physics, Department of Physics, Universit\"at Erlangen-N\"urnberg, 91058 Erlangen}
\author{Aashish A. Clerk}
\affiliation{Department of Physics, McGill University, 3600 rue University, Montr\'{e}al, Quebec H3A 2T8, Canada}
\author{Oskar Painter}
\affiliation{Kavli Nanoscience Institute, California Institute of Technology, Pasadena, California 91125, USA}
\affiliation{Institute for Quantum Information and Matter and Thomas J. Watson, Sr., Laboratory of Applied Physics, California Institute of Technology, Pasadena, California 91125, USA}
\email{opainter@caltech.edu}

\begin{abstract} 

  Synthetic magnetism has been used to control charge neutral excitations for applications ranging from classical beam steering to quantum simulation. In optomechanics, radiation-pressure-induced parametric coupling between optical (photon) and mechanical (phonon) excitations may be used to break time-reversal symmetry, providing the prerequisite for synthetic magnetism. Here we design and fabricate a silicon optomechanical circuit with both optical and mechanical connectivity between two optomechanical cavities. Driving the two cavities with phase-correlated laser light results in a synthetic magnetic flux, which in combination with dissipative coupling to the mechanical bath, leads to nonreciprocal transport of photons with $35$~dB of isolation.  Additionally, optical pumping with blue-detuned light manifests as a particle non-conserving interaction between photons and phonons, resulting in directional optical amplification of $12$~dB in the isolator through direction. These results indicate the feasibility of utilizing optomechanical circuits to create a more general class of nonreciprocal optical devices, and further, to enable novel topological phases for both light and sound on a microchip.

\end{abstract}

\maketitle

\begin{figure}[!htb]
\begin{center}
\includegraphics[width=\columnwidth]{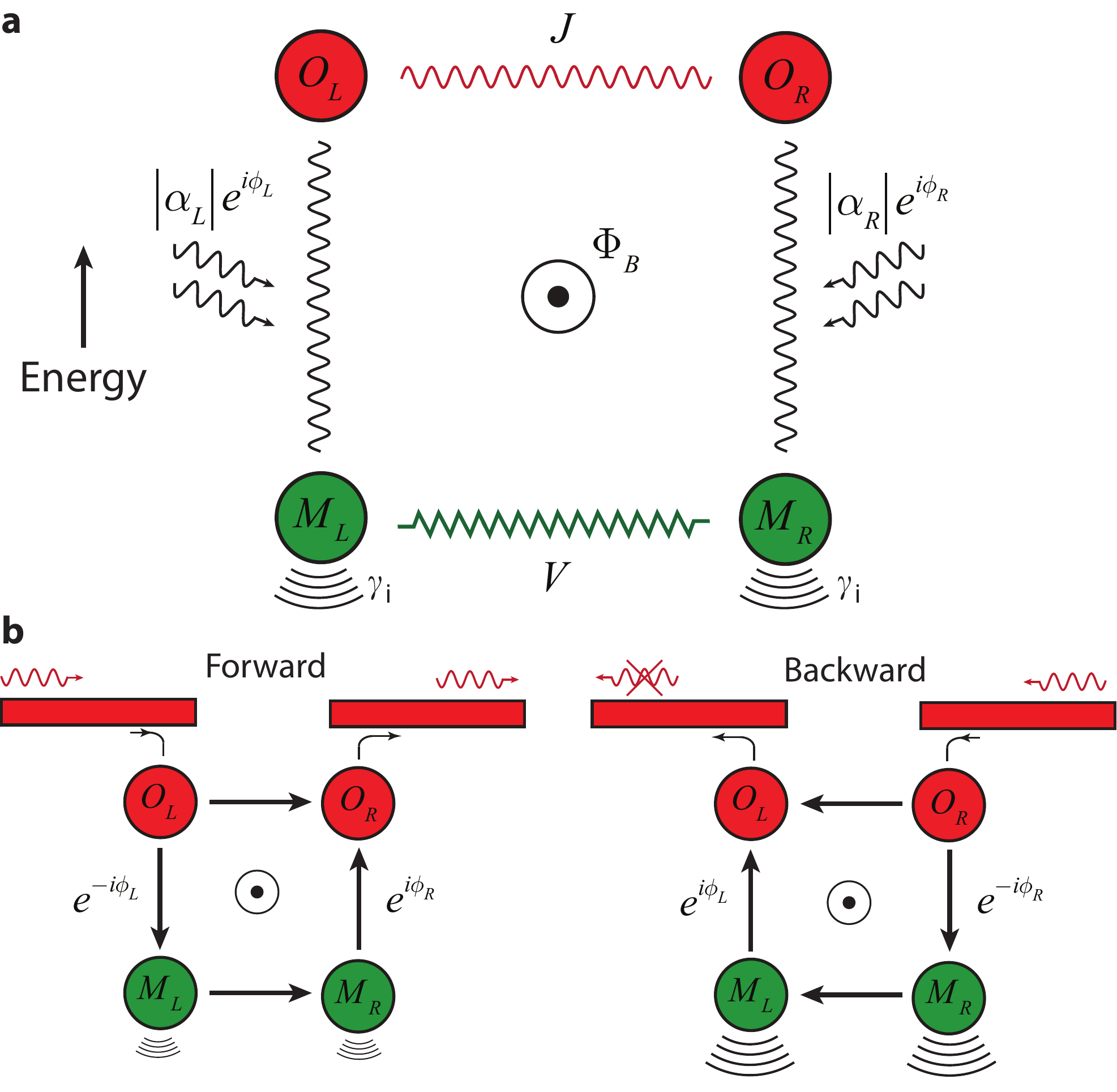}
\caption{\textbf{Synthetic magnetic field in an optomechanical cavity system.} \textbf{a}, In this scheme consisting of only two optomechanical cavities, a two-dimensional plaquette can be formed from the synthetic dimension~\cite{manybody2} created by radiation pressure coupling from the optical modes to the mechanical modes.  Photon hopping at rate $J$ and phonon hopping at rate $V$ occurs between the optical and mechanical cavities, respectively, with $J$ and $V$ real for appropriate choice of gauge.  Pumping of the optomechancial cavities with phase correlated laser light ($|\alpha_{\text{L}}|e^{i\phi_{\text{L}}}$ for the left cavity and ($|\alpha_{\text{R}}|e^{i\phi_{\text{R}}}$ for the right cavity) results in a synthetic flux $\PhiB = \phi_{\text{L}} - \phi_{\text{R}}$ threading the 4-mode plaquette. \textbf{b}, Scheme for detecting the synthetic flux through nonreciprocal power transmission of an optical probe laser field.  For forward ($\text{L} \rightarrow \text{R}$) propagation, constructive interference set by the flux-dependent phase $\PhiB \approx \pi/2$ of the dissipative phonon coupling path results in efficient optical power transmission.  The accumulated phase in the phonon coupling path is reversed for the backward ($\text{R} \rightarrow \text{L}$) propagation direction resulting in destructive interference and reduced optical power transmission in the left output waveguide.  The power in this case is sunk into the mechanical baths.  
}
\label{fig1}
\end{center}
\end{figure}

Synthetic magnetism involving charge neutral elements such as atoms~\cite{Dalibard2011}, polaritons~\cite{Koch2010,Devoret2015,Martinis2016}, and photons~\cite{Umucal2011,Hafezi2011,fang,Rechtsman2013,Tzuang2014} is an area of active theoretical and experimental research, driven by the potential to simulate quantum many-body phenomena~\cite{Lin2009}, reveal new topological wave effects~\cite{Ray2014,Lu2014}, and create defect-immune devices for information communication~\cite{Hafezi2011,Tzuang2014}.  Optomechanical systems~\cite{RMP}, involving the coupling of light intensity to mechanical motion via radiation pressure, are a particularly promising venue for studying synthetic fields, as they can be used to create the requisite large optical nonlinearities~\cite{Rosenberg2009}.  By applying external optical driving fields time-reversal symmetry may be explicitly broken in these systems. It was predicted that this could enable optically tunable nonreciprocal propagation in few-port devices~\cite{Manipatruni2009,Hafezi2012,ophwg2,Wang2015}, or in the case of a lattice of optomechanical cavities, topological phases of light and sound~\cite{manybody1,manybody2}. Here we demonstrate a generalized form of optical nonreciprocity in a silicon optomechanical crystal circuit~\cite{omcc} that goes beyond simple directional propagation; this is achieved using a combination of synthetic magnetism, reservoir engineering, and parametric squeezing.


Distinct from recent demonstrations of optomechanical nonreciprocity in degenerate whispering-gallery resonators with inherent nontrivial topology~\cite{Kim2015,Dong2016,Ruesink2016}, we employ a scheme similar to that proposed in Refs.~\cite{ophwg2,manybody2} in which a synthetic magnetic field is generated via optical pumping of the effective lattice formed by coupled optomechanical cavities.  In such a scenario, the resulting synthetic field amplitude is set by the spatial variation of the pump field phase and the field lines thread optomechanical plaquettes between the photon and phonon lattices (see Fig.~\ref{fig1}).  To achieve nonreciprocal transmission of intensity in the two-port device of this work -- i.e., bonafide phonon or photon transport effects, not just nonreciprocal transmission phase -- one can combine this synthetic field with dissipation to implement the general reservoir engineering strategy outlined in Ref.~\cite{reservoir}.  This approach requires one to balance coherent and dissipative couplings between optical cavities.  In our system the combination of the optical drives and mechanical dissipation provide the ``engineered reservoir'' which is needed to mediate the required dissipative coupling.

To highlight the flexibility of our approach, we use it to implement a novel kind of nonreciprocal device exhibiting gain~\cite{Abdo2013,Abdo2014}.  By using an optical pump which is tuned to the upper motional sideband of the optical cavities, we realize a two-mode squeezing interaction which creates and destroys photon and phonon excitations in pairs. These particle non-conserving interactions can be used to break time-reversal symmetry in a manner that is distinct from a standard synthetic gauge field.  In a lattice system, this can enable unusual topological phases and surprising behavior such as protected chiral edge states involving inelastic scattering~\cite{Peano2016} and amplification~\cite{Peano2016B}.  Here, we use these interactions along with our reservoir-engineering approach to create a cavity-based optical directional amplifier:  backward propagating signals and noise are extinguished by $35$~dB relative to forward propagating waves which are amplified with an internal gain of $12$~dB ($1$~dB port-to-port).   

The optomechanical system considered in this work is shown schematically in Fig.~\ref{fig1}a and consists of two interacting optomechanical cavities, labeled $L$ (left) and $R$ (right), with each cavity supporting one optical mode $\OLR$ and one mechanical mode $\MLR$.  Both the optical and mechanical modes of each cavity are coupled together via a photon-phonon waveguide, resulting in optical and mechanical inter-cavity hopping rates of $J$ and $V$, respectively (here we choose a local definition of the cavity amplitudes so both are real). The radiation pressure interaction between the co-localized optical and mechanical modes of a single cavity can be described by a Hamiltonian $\hat{\mathcal{H}}=\hbar \gzeroO \adag \ahat(\bhat+\bdag)$, where $\ahat(\bhat)$ is the annihilation operator of the optical (mechanical) mode and $\gzeroO$ is the vacuum optomechanical coupling rate~\cite{RMP} (here we have omitted the cavity labeling). 

To enhance the effective photon-phonon interaction strength each cavity is driven by an optical pump field with frequency relatively detuned from the optical cavity resonance by the mechanical frequency ($\Delta \equiv \omegapO - \omegacO \approx \pm\omegamO$), with a resulting intra-cavity optical field amplitude $|\alpha|e^{i\phi}$.   In the good-cavity limit, where $\omegamO \gg \kappa$ ($\kappa$ being the optical cavity linewidth), spectral filtering by the optical cavity preferentially selects resonant photon-phonon scattering, leading to a linearized Hamiltonian with either a two-mode squeezing form $\hat{\mathcal{H}}_{\text{ent}}=\hbar G(e^{i\phi} \ddagg \bdag + e^{-i\phi} \dhat\bhat)$ (blue detuned pumping) or a beamsplitter form $\hat{\mathcal{H}}_{\text{ex}}=\hbar G(e^{i\phi} \ddagg\bhat + e^{-i\phi} \dhat\bdag)$ (red detuned pumping).  Here $G=\gzeroO|\alpha|$ is the parametrically enhanced optomechanical coupling rate and $\dhat = \ahat - \alpha$ contains the small signal sidebands of the pump.  For both cases the phase of the resulting coupling coefficient is nonreciprocal in terms of the generation and annihilation of photon-phonon excitations. As has been pointed out before, such a nonreciprocal phase resembles the Peierls phase that a charged particle accumulates in a magnetic vector potential~\cite{ab}. Crucially, the relative phase $\PhiB=\phiL-\phiR$ is gauge independent (i.e.~independent of local redefinitions of the $\ahat$ and $\bhat$ cavity amplitudes), implying it should have an observable effect.  In the simple case of $\Delta = -\omegamO$, $\PhiB$ is formally equivalent to having a synthetic magnetic flux threading the plaquette formed by the four coupled optomechanical modes (two optical and two mechanical)\cite{fang,ophwg2,manybody2}.  For $\Delta = +\omegamO$, a non-zero $\PhiB$ still results in the breaking of time-reversal symmetry, though the lack of particle number conservation means that it is not simply equivalent to a synthetic gauge field.  Nonetheless, we will refer to it as a flux in what follows for simplicity.  

\begin{figure*}[t!]
\begin{center}
\includegraphics[width=2\columnwidth]{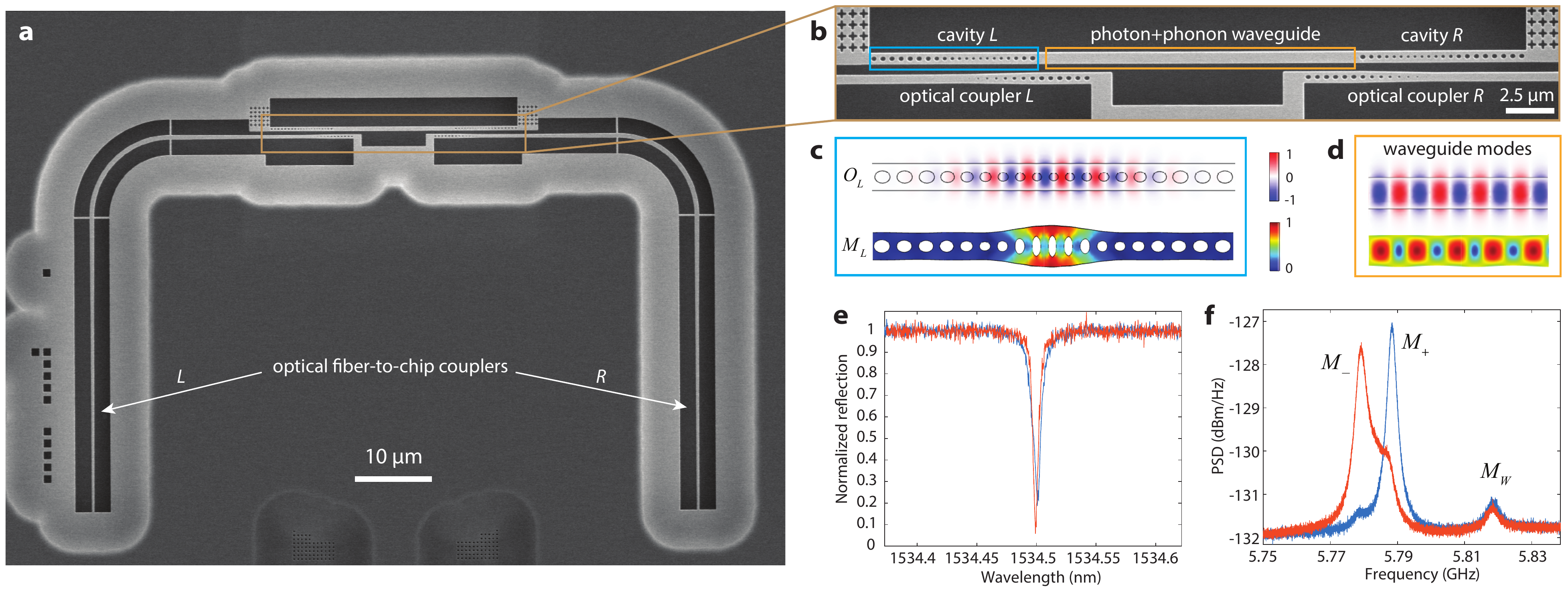}
\caption{ \textbf{Silicon optomechanical crystal circuit.} \textbf{a}, Scanning electron microscopy (SEM) image of the optomechanical crystal circuit studied in this work.  The circuit is fabricated from a silicon-on-insulator microchip (see App.~\ref{App:A}). \textbf{b}, SEM of the main part of the circuit, which consists of a left and a right nanobeam optomechanical crystal cavity with a central unpatterned nanobeam waveguide connecting the two cavities.  A left and right optical coupler, which are each fed by an adiabatic fiber-to-chip coupler~\cite{Groeblacher2013}, are used to evanescently couple light into either of the two optical cavities.  \textbf{c}, FEM simulated electrical field $E_{y}$ and magnitude of the displacement field for the localized optical and mechanical cavity modes, respectively, of the nanobeam.  \textbf{d}, FEM simulated section of the corresponding optical and mechanical modes of the connecting waveguide.  \textbf{e,} Optical reflection spectrum of the left (blue) and right (orange) optical cavities. \textbf{f}, Optically transduced mechanical power spectral density (PSD) measured from the left (blue) and right (orange) optical cavities. $\Mpm$ are the two hybridized mechanical cavity modes with frequency $\omega_{M_{+(-)}}/2\pi=5788.4$ $(5779.1)~$MHz and $\Mw$ is a mechanical waveguide mode with frequency $\omegamW/2\pi=5818.3$~MHz.} 
\label{fig2}
\end{center}
\end{figure*}


To detect the presence of the effective flux $\PhiB$, consider the transmission of an optical probe signal, on resonance with the optical cavity resonances and coupled in from either the left or the right side via external optical coupling waveguides as depicted in Fig.~\ref{fig1}b. The probe light can propagate via two different paths simultaneously: (i) direct photon hopping between cavities via the connecting optical waveguide, and (ii) photon-phonon conversion in conjunction with intervening phonon hopping via the mechanical waveguide between the cavities.  As in the Aharonov-Bohm effect for electrons~\cite{Aharanov1959}, the synthetic magnetic flux set up by the phase-correlated optical pump beams in the two cavities causes a flux-dependent interference between the two paths. 
We define the forward (backward) transmission amplitude as $\TLR \equiv d_{\text{out,L(R)}}/d_{\text{in,R(L)}}$, where $d_{\text{out}(\text{in})}$ is the amplitude of the outgoing (incoming) electromagnetic signal field in the corresponding coupling waveguide in units of square root of photon flux.
The optical transmission amplitude in the forward direction has the general form 


\be 
\TR [\omega;\Delta=\pm \omegamO] = \Apm [\omega]\left(J -\Gammapm [\omega]e^{- i\PhiB}\right), \label{eq:TR}
\ee

\noindent where $\omega \equiv \omegas-\omegap$ and $\omegas$ is the frequency of the probe light. $\Gammapm$ is the amplitude of the effective mechanically-mediated coupling between the two optical cavities, and is given by

\be\label{eq1b}
\Gammapm [\omega] = \frac{V \GL \GR}{( -i(\omega \pm \omegamL)+ \frac{\gammaiL}{2})( -i(\omega \pm \omegamR)+ \frac{\gammaiR}{2})+V^2}.
\ee

\noindent The prefactor $\Apm [\omega]$ in Eq.~(\ref{eq:TR}) accounts for reflection and loss at the optical cavity couplers, as well as the mechanically-induced back-action on the optical cavities.  This prefactor is independent of the transmission direction, and for the reverse transmission amplitude $\TL$, only the sign in front of $\PhiB$ changes. 

The directional nature of the optical probe transmission may be studied via the frequency-dependent ratio

\be\label{eq1}
\left(\frac{\TR}{\TL}\right)[\omega; \Delta=\pm\omegamO] = \frac{J-\Gammapm[\omega]e^{- i\Phi_B}}{J - \Gammapm[\omega]e^{+ i\Phi_B}}.
\ee

\noindent Although the presence of the synthetic flux breaks time-reversal symmetry, it does not in and of itself result in nonreciprocal photon transmission magnitudes upon swapping input and output ports~\cite{Deak2012,reservoir}.  In our system, if one takes the limit of zero intrinsic mechanical damping (i.e.~$\gammaik=0$), the mechanically-mediated coupling amplitude $\Gammapm [\omega]$ is real at all frequencies.  This implies $|\TR| = |\TL|$, irrespective of the value of $\PhiB$.  We thus find that non-zero mechanical dissipation will be crucial in achieving any non-reciprocity in the magnitude of the optical transmission amplitudes.

The general reservoir-engineering approach to nonreciprocity introduced in Ref.~\cite{reservoir} provides a framework for both understanding and exploiting the above observation.  It demonstrates that nonreciprocity is generically achieved by balancing a direct (Hamiltonian) coupling between two cavities against a dissipative coupling of the cavities; such a dissipative coupling can arise when both cavities couple to the same dissipative reservoir.  The balancing requires both a tuning of the magnitude of the coupling to the bath, as well as a relative phase which plays a role akin to the flux $\PhiB$.  In our case, the damped mechanical modes can play the role of the needed reservoir, with the optical drives controlling how the optical cavities couple to this effective reservoir.  One finds that at any given frequency $\omega$, the mechanical modes induce both an additional coherent coupling between the two cavities (equivalent to an additional coupling term in the Hamiltonian) as well as a dissipative coupling (which is not describable by a Hamiltonian).  As is shown explicitly in App,~\ref{App:B}, in the present setting these correspond directly to the real and imaginary parts of $\Gammapm[\omega]$.  Hence, the requirement of having $\textrm{Im }\Gamma[\omega] \neq 0$ is equivalent to requiring a non-zero mechanically-mediated dissipative coupling between the cavities.

Achieving directionality requires working at a frequency where the dissipative coupling has the correct magnitude to balance the coherent coupling $J$, and a tuning of the flux $\PhiB$.  For $|\Gammapm [\omega]| = J$ and $\arg(\Gammapm) = -\PhiB$ ($\neq 0, \pi$), one obtains purely uni-directional transport where the right optical cavity is driven by the left optical cavity but not vice versa. One finds from Eq.~(\ref{eq1}) that the mechanically-mediated dissipative coupling between the cavities is maximized at frequencies near the mechanical normal mode frequencies $\omega \approx -\omegamO \pm V$; to achieve the correct magnitude of coupling, the optical pumping needs to realize a many-photon optomechanical coupling $\Gk \approx (J \gammaik)^{1/2}$ (see App.~\ref{App:B} for details).  Note that our discussion applies to both the choices of red-detuned and blue-detuned pumping.  While the basic recipe for directionality is the same, in the blue-detuned pump case the effective reservoir seen by the cavity modes can give rise to negative damping, with the result that the forward transmission magnitude can be larger than one.  We explore this more in what follows.

\begin{figure*}[t!]
\begin{center}
\includegraphics[width=2\columnwidth]{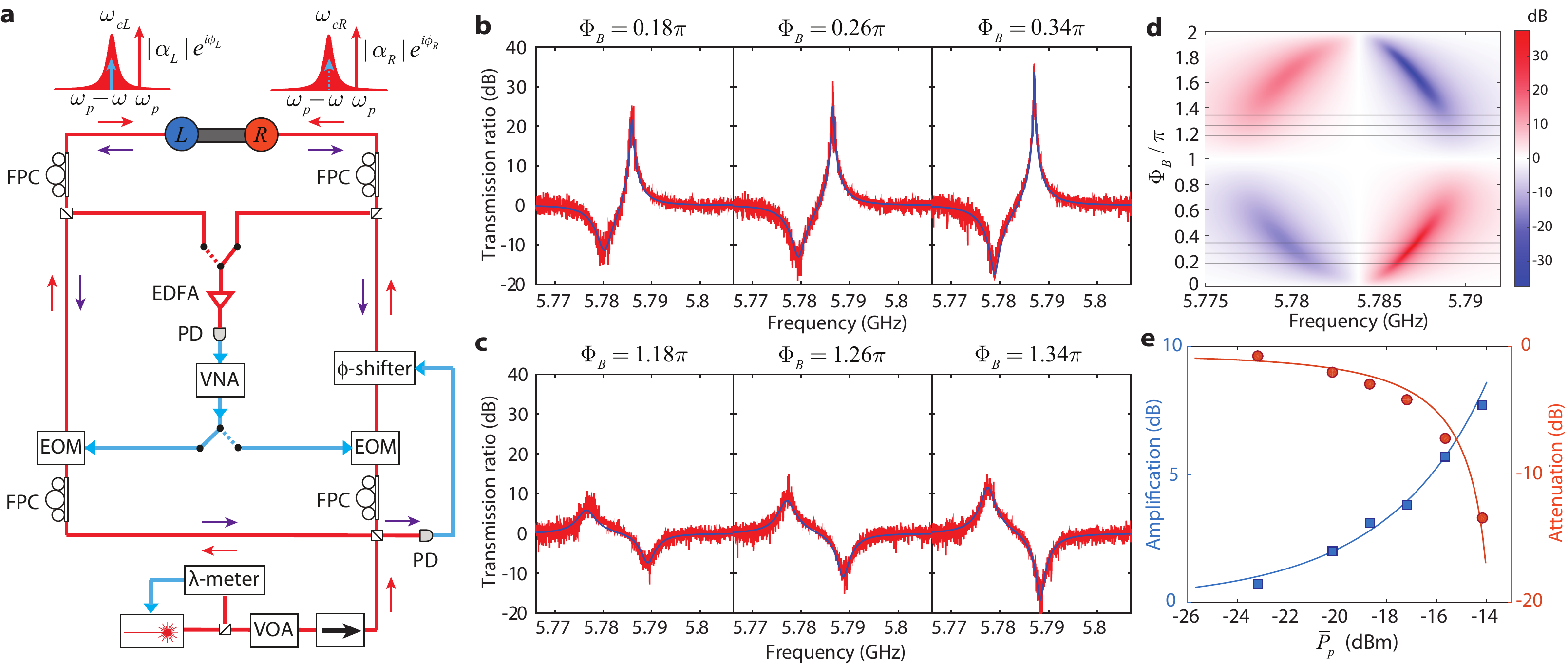}
\caption{ \textbf{Measurement of optical nonreciprocity.} \textbf{a}, Experiment set-up. Red (blue) lines are optical (electronic) wiring. Blue-detuned pump light from a tunable diode laser is split into two paths and fed into the two cavities (red arrows). Part of the reflected pump laser light from the cavities (purple arrows) is collected by a photodetector (PD) and fed into a stretchable fiber phase shifter ($\phi$-shifter) to tune and lock the phase difference of the optical pumps. Each optical path can be modulated by an electro-optic modulator (EOM) to generate an optical sideband which we use as the optical probe signal. The microwave modulation signal with frequency $\omegamod$ is generated by port 1 of a vector network analyzer (VNA).  After optical amplification and photodetection, the transmitted optical probe signal through the optomechanical circuit is sent back to port 2 of the VNA to measure the phase and amplitude of the optical probe transmission coefficient. EDFA: Erbium doped fiber amplifier, FPC: fiber polarization controller, $\lambda$-meter: wavelength meter. \textbf{b}, The ratio of optical power transmission coefficients for right- and left-propagation versus modulation frequency ($\omegamod = -\omega = \omegap-\omegas$), for three different synthetic flux values $\PhiB/\pi=0.18$, $0.26$, and $0.34$. The blue curves  correspond to the fit of the theoretical model (c.f. Eq.~\ref{eq1}) to the measured spectra. \textbf{c}, The power transmission coefficient ratio for $\PhiB$ with an additional $\pi$ flux relative to those in \textbf{b}. \textbf{d}, Theoretical calculation of the power transmission coefficient ratio for $0\le\PhiB\le 2\pi$, where the six grey lines correspond to the six measured $\Phi_B$ values in \textbf{b} and \textbf{c}. \textbf{e}, Peak forward signal amplification above background level (blue squares) and corresponding signal attenuation in the reverse direction (red circles) versus average optical pump power ($\Ppbar =\sqrt{\PpL\PpR}$) for fixed flux value of $\PhiB=0.28\pi$. The solid curves are theoretical calculations based upon the theoretical model (c.f. Eq.~\ref{eq1} and SI) fit to the data in \textbf{b} and \textbf{c}.}
\label{fig3}
\end{center}
\end{figure*}

In order to realize the optomechanical circuit depicted in Fig.~\ref{fig1} we employ the device architecture of optomechanical crystals~\cite{omc,phoxonic2,phoxonic3}, which allows for the realization of integrated cavity-optomechanical circuits with versatile connectivity and cavity coupling rates~\cite{shielding,omcc}. Figure~\ref{fig2}a shows the optomechanical crystal circuit fabricated on a silicon-on-insulator microchip. The main section of the circuit, shown zoomed-in in Fig.~\ref{fig2}b, contains two optomechanical crystal nanobeam cavities, each of which has an optical resonance of wavelength $\lambda\approx 1530$~nm and a mechanical resonance of frequency $\omegamO/2\pi \approx 6$~GHz.  The two optical cavities can be excited through two separate optical coupling paths, one for coupling to the left cavity and one for the right cavity.  Both the left and right optical coupling paths consist of an adiabatic fiber-to-chip coupler which couples light from an optical fiber to a silicon waveguide, and a near-field waveguide-to-cavity reflective coupler.  This allows separate optical pumping of each cavity and optical transmission measurements to be carried out in either direction.  The two nanobeam cavities are physically connected together via a central silicon beam section which is designed to act as both an optical waveguide and an acoustic waveguide.  The central beam thus mediates both photon hopping and phonon hopping between the two cavities even though the cavities are separated by a distance much larger than the cavity mode size~\cite{noda,omcc}.  The numerically simulated mode profiles for the localized cavities and the connecting waveguide are shown in Fig.~\ref{fig2}c and \ref{fig2}d, respectively.  The hopping rate for photons and phonons can be engineered by adjusting the number and shape of the holes in the mirror section of the optomechanical crystal cavity along with the free-spectral range of the connecting waveguide section~\cite{omcc}.  Here we aim for a design with $J/2\pi \approx 100$~MHz and $V/2\pi \approx 3$~MHz so that nonreciprocity can be realized at low optical pump power, yet still with high transmission efficiency.

As will be presented elsewhere~\cite{afm}, the optical and mechanical frequencies of the optomechanical cavities are independently trimmed into alignment post-fabrication using an atomic force microscope to oxidize nanoscale regions of the cavity.  After nano-oxidation tuning, the left (right) cavity has optical resonance wavelength $\lambdaLR=1534.502$ $(1534.499)$~nm, total loaded damping rate $\kappaLR/2\pi=1.03$ $(0.75)$~GHz, and intrinsic cavity damping rate $\kappaiLR/2\pi=0.29$ $(0.31)$~GHz (c.f. Fig.~\ref{fig2}e).  Note that hybridization of the optical cavity resonances is too weak to be observable in the measured left and right cavity spectra due to the fact that the optical cavity linewidths are much larger than the designed cavity coupling $J$.  The thermal mechanical spectra, as measured from the two cavities using a blue-detuned pump laser (see App.~\ref{App:A}), are shown in Fig.~\ref{fig2}f where one can see hybridized resonances $\Mpm$ which are mixtures of the localized mechanical cavity modes $\ML$ and $\MR$.  A nearby phonon waveguide mode ($\Mw$) is also observable in both left and right cavity spectra.  The optomechanical coupling rate and mechanical dissipation rate of $\MLR$ were measured before nano-oxidation tuning, yielding $\gzeroLR/2\pi=0.76$ $(0.84)$~MHz and $\gammaiLR/2\pi=4.3$ $(5.9)$~MHz.

The experimental apparatus used to drive and probe the optomechanical circuit is shown schematically in Fig.~\ref{fig3}a.  As indicated, an optical pump field for the left and right cavities is generated from a common diode laser.  The phase difference of the pump fields at the input to the cavities, and thus the synthetic magnetic flux, is tuned by a stretchable fiber phase shifter and stabilized by locking the interference intensity of the reflected pump signals from the cavities.  To highlight the unique kinds of nonreciprocal transport possible in our setup, we present results for an experiment performed with   
blue-detuned pump fields with frequency $\omegap\approx\omegacO+\omegamO$; as discussed, this will enable non-reciprocal transport with gain.  An input optical probe signal is generated from either of the left or right cavity pump beams by sending them through an electro-optic modulator (EOM).  A vector network analyzer (VNA) is used to drive the EOMs at modulation frequency $\omegamod$ and detect the photocurrent generated by the beating of the transmitted probe and reflected pump signals, thus providing amplitude and phase information of the transmitted probe signal.  Owing to the spectral filtering of the cavities, only the generated lower sideband of the blue-detuned pump at $\omega = -\omegamod$ is transmitted through the circuit as a probe signal.  

\begin{figure}[t!]
\begin{center}
\includegraphics[width=\columnwidth]{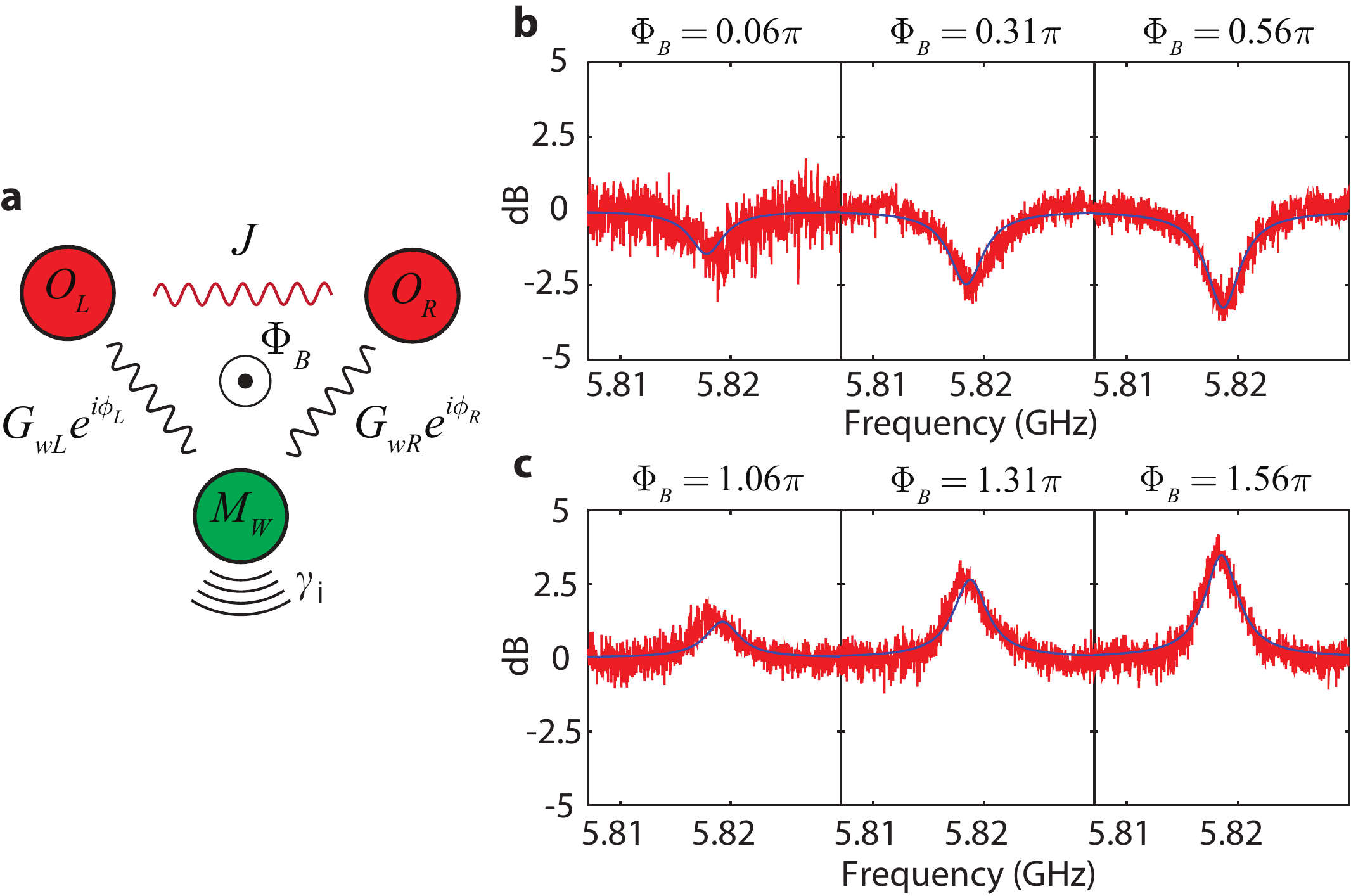}
\caption{ \textbf{Synthetic magnetic field with a single mechanical cavity.} \textbf{a}, Physical configuration for generation of a synthetic magnetic field and optical nonreciprocity with two optical modes parametrically coupled with a common dissipative mechanical waveguide mode. \textbf{b,c} The ratio of optical power transmission coefficients for right and left propagation versus modulation frequency $\omegamod$ around the frequency of the waveguide mode $\Mw$ for various $\PhiB$. The blue curves correspond to a fit of the theoretical model (see App.~\ref{App:B}) to the measured data.}
\label{fig4}
\end{center}
\end{figure}

Figure~\ref{fig3}b shows the ratio of the forward and backward optical power transmission coefficients of the probe light ($|\TR/\TL|^2$) for several magnetic flux values between $\PhiB = 0$ and $\pi$.  For these measurements the pump powers at the input to the left and right cavity were set to $\PpL=-14.2$~dBm and $\PpR=-10.8$~dBm, respectively, corresponding to intra-cavity photon numbers of $\ncL=1000$ and $\ncR=1420$.  So as to remove differences in the forward and reverse transmission paths external to the optomechanical circuit, here the $|\TR/\TL|^2$ ratio is normalized to $0$~dB for a modulation frequency $\omegamod/2\pi \approx 5.74$~GHz, detuned far from mechanical resonance in a frequency range where reciprocal transmission is expected. Closer to mechanical resonance, strong nonreciprocity in the optically transmitted power is observed, with a peak and a dip in $|\TR/\TL|^2$ occurring roughly at the resonance frequencies of the hybridized mechanical modes $M_{+}$ and $M_{-}$, respectively (c.f. Fig.~\ref{fig2}c).  The maximum contrast ratio between forward and backward probe transmission -- the isolation level -- is measured to be $35$~dB for $\PhiB=0.34\pi$ near the $M_{+}$ resonance.  The forward transmission is also amplified in this configuration (blue-detuned pump, $\Delta=+\omegamO$), with a measured peak probe signal amplification of $12$~dB above the background level set by photon hopping alone ($J/|\Gammapm| \gg 1$).  The corresponding port-to-port net gain is only $1$~dB due to impedance mismatching ($J\neq\kappa/2$) and intrinsic optical cavity losses (see SI for details).

From a two-parameter fit to the measured optical power transmission ratio spectra using Eq.~\ref{eq1} (see blue curves in Figs.~\ref{fig3}b and \ref{fig3}c), we obtain a waveguide-mediated optical and mechanical hopping rate of $J/2\pi=110$~MHz and $V/2\pi=2.8$~MHz, respectively,  consistent with our design parameters. Figure~\ref{fig3}d shows the theoretical calculation of $|\TR/\TL|^2$ for a full $2\pi$ range of $\PhiB$ with the measured and fit optomechanical circuit parameters. The pattern is seen to be odd symmetric with respect to $\PhiB=\pi$.  Inserting an additional magnetic flux $\pi$ into the measurements performed in Fig.~\ref{fig3}b yields the spectra shown in Fig.~\ref{fig3}c which displays a switch in the isolation direction as predicted by the model.  The pump power dependence of the peak (in frequency) forward signal amplification and the corresponding backward signal attenuation relative to the background level far from mechanical resonance are shown in Fig.~\ref{fig3}e for a fixed magnetic flux of $\PhiB=0.28\pi$.  Good correspondence with the theoretical power dependence (solid curves) is observed, with nonreciprocal amplification vanishing at low pump power.  

One can also obtain nonreciprocal optical power transmission utilizing an even simpler system involving a single mechanical cavity.  This is the situation we have for the Fabry-Perot-like mechanical resonances that exist in the central coupling waveguide (see $\Mw$ resonance of Fig.~\ref{fig2}c).  As depicted in Fig.~\ref{fig4}a, the mode configuration in this case consists of two optical cavity modes ($\OL$ and $\OR$) coupled together via the optical waveguide, one mechanical waveguide mode $\Mw$ which is parametrically coupled to each of the optical cavity modes, and the synthetic magnetic flux $\PhiB=\phiL-\phiR$ due to the relative phases of the optical pump fields threading the triangular mode space. In Fig.~\ref{fig4}b and \ref{fig4}c we show the measurement of $|\TR/\TL|^2$ for a series of different flux values $\PhiB$ with blue-detuned pumping ($\Delta\approx +\omegamW$) at levels of $\ncL=770$ and $\ncR=1090$.  In this single mechanical mode case the direction of the signal propagation is determined by the magnitude of the flux; $\PhiB \le \pi$ leads to backward propagation and $\PhiB \ge \pi$ to forward propagation.  The lower contrast ratio observed is a result of the weaker coupling between the localized optical cavity modes and the external waveguide mode, which for the modest pump power levels used here ($\lesssim 100$~$\mu$W) does not allow us to reach the parametric coupling required for strong directional transmission.





While our focus has been on the propagation of injected coherent signals through the optomechanical circuit, it is also interesting to consider the flow of noise.  As might be expected, the induced directionality of our system also applies to noise photons generated by the upconversion of both thermal and quantum fluctuations of the mechanics; see App.~\ref{App:C} for detailed calculations.  One finds that for the system of Fig.~\ref{fig2}, the spectrally-resolved photon noise flux shows high directionality, but that the sign of this directionality changes as a function of frequency (analogous to what happens in the transmission amplitudes).  In contrast, in the single-mechanical mode setup of Fig.~\ref{fig4} the sign of the directionality is constant with frequency, and thus the total (frequency-integrated) noise photon flux is directional depending upon the flux magnitude. The laser pump fields can thus effectively act as a heat pump, creating a temperature difference between the left and right waveguide output fields.  The corresponding directional flow of quantum noise is especially useful for quantum information applications, as it can suppress noise-induced damage of a delicate signal source like a qubit~\cite{reservoir,Abdo2014}. 

The device studied in this work highlights the potential for optomechanics to realize synthetic gauge fields and novel forms of nonreciprocity enabled by harnessing mechanical dissipation.  Using just a few modes, it was possible to go beyond simply mimicking the physics of an isolator and realize a directional optical amplifier.  By adding modes, an even greater variety of behaviours could be achieved.   For example, the simple addition of a third optical cavity mode, tunnel-coupled to the first two cavities but with no mechanical coupling, would realize a photon circulator similar to the phonon circulators considered in Ref.~\cite{ophwg2}.  Scaling the synthetic gauge field mechanism realized here to a full lattice of optomechanical cavities would allow the study of topological phenomena in the propgation of both light and sound.  Predicted effects include the formation of back-scattering immune photonic~\cite{manybody2} and phononic~\cite{manybody1} chiral edge states, topologically nontrivial phases of hybrid photon-phonon excitations~\cite{manybody1}, dynamical gauge fields~\cite{dynamicalgauge}, and, in the case of non-particle-conserving interactions enabled by blue-detuned optical pumping, topologically protected inelastic scattering of photons~\cite{Peano2016} and even protected amplifying edge states~\cite{Peano2016B}.



\begin{acknowledgments} 
The authors would like to thank Michael Roukes for the use of his atomic force microscope in the nanooxidation tuning of the cavities.  This work was supported by the AFOSR-MURI Quantum Photonic Matter, the ARO-MURI Quantum Opto-Mechanics with Atoms and Nanostructured Diamond (grant N00014-15- 1-2761), the University of Chicago Quantum Engineering Program (AAC,AM), the ERC Starting Grant OPTOMECH (FM), the Institute for Quantum Information and Matter, an NSF Physics Frontiers Center with support of the Gordon and Betty Moore Foundation, and the Kavli Nanoscience Institute at Caltech.
\end{acknowledgments}



\onecolumngrid
\appendix


\section{Device Fabrication and Methods}
\label{App:A}

\subsection{Device fabrication and atomic force microscope nano-oxidation tuning}
The devices were fabricated from a silicon-on-insulator wafer with a silicon device layer thickness of $220$~nm and buried-oxide layer thickness of $2$~$\mu$m. The device geometry was defined by electron-beam lithography followed by inductively coupled plasma reactive ion etching to transfer the pattern through the $220$~nm silicon device layer. The devices were then undercut using an HF:H$_2$O solution to remove the buried oxide layer and cleaned using a piranha etch. 

After device fabrication, we used an atomic force microscope to draw nanoscale oxide patterns on the silicon device surface.  This process modifies the optical and mechanical cavity frequencies in a controllable and independent way with the appropriate choice of oxide pattern. The nano-oxidation process was carried out using an Asylum MFP-3D atomic force microscope and conductive diamond tips (NaDiaProbes) in an environment with relative humidity of $48\%$. The tip was biased at a voltage of $-11.5$~V, scanned with a velocity of $100$~nm/s, and run in tapping mode with an amplitude of $10$~nm.  The unpassivated silicon device surface was grounded.

\subsection{Optical transmission coefficient measurement}

\begin{figure}[htp!]
\begin{center}
\includegraphics[width=0.7\columnwidth]{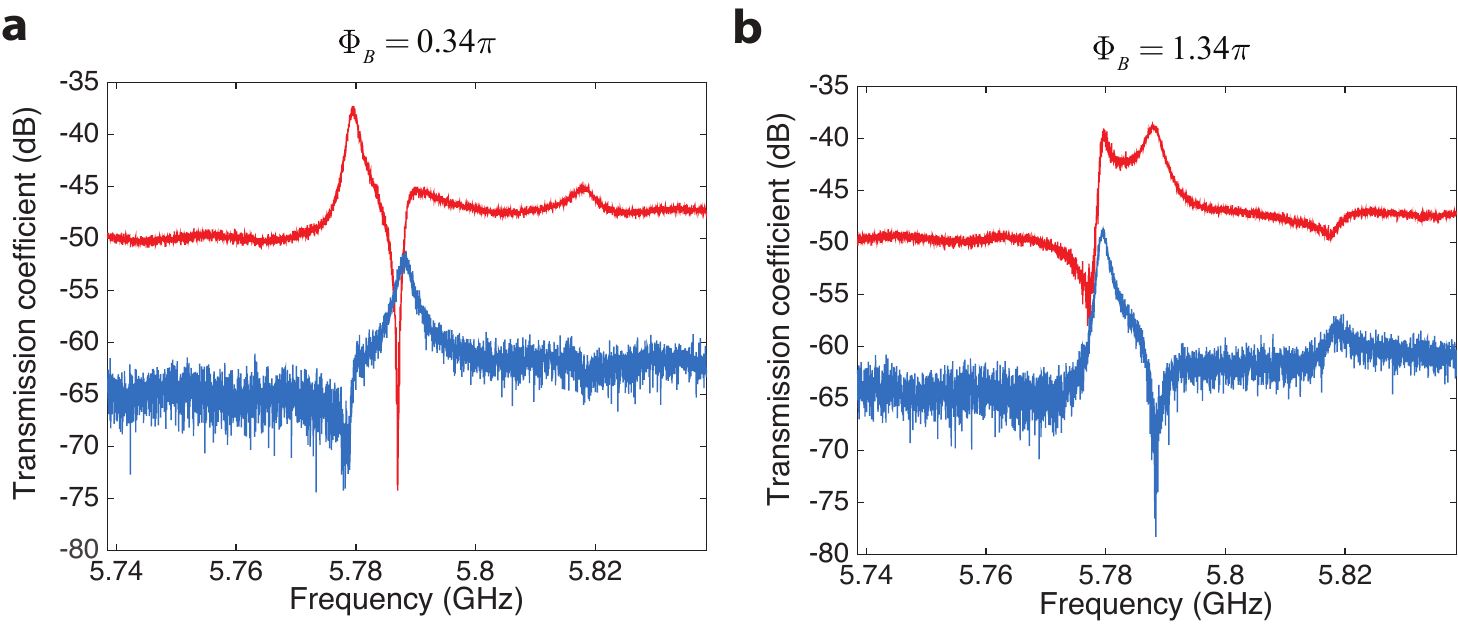}
\caption{ \textbf{a} Microwave signal power transmission through the optomechanical circuit for forward (right-propagation; blue) and backward (left-propagation; blue curve) directions, with flux set to $\PhiB=0.34\pi$ and cavity photon number $n_{cL}=1000$ and $n_{cR}=1420$. \textbf{b} Same as \textbf{a} but with $\PhiB=1.34\pi$.}
\label{figs1}
\end{center}
\end{figure}

To measure the optical power transmission through the optomechanical circuit we used a vector network analyzer (VNA). The VNA outputs a microwave tone from port 1 with frequency $\omegamod$ to an electro-optic modulator which modulates the optical pump to generate an optical sideband corresponding to the optical probe. In the case of a blue-detuned pump from the optical cavity resonance, the probe field corresponds to the lower sideband (selected by the filtering properties of the cavity itself).  Both the optical probe and pump are launched into one optomechanical cavity in the circuit. At the other cavity, the transmitted optical probe combines with a second pump and the beating of the two is detected by a high-speed photodetector (both the first and second pump beams are from the same laser source, and thus phase coherent). The photocurrent signal from the photodetector is sent into port 2 of the VNA to measure the microwave signal transmission coefficient $T_\mu$. Fig.~\ref{figs1} shows $|T_\mu|^2$ for forward (right-propagating; blue curve) and backward (left-propagating; red curve) directions through the optomechanical circuit as a function of the modulation frequency $\omegamod$.  In Fig.~\ref{figs1}a the synthetic flux value is locked to $\PhiB=0.34\pi$ whereas in Fig.~\ref{figs1}b $\PhiB=1.34\pi$.  In both flux settings the optical pumping levels were such that the left and right cavity photon numbers were $n_{cL}=1000$ and $n_{cR}=1420$, respectively.  This is the raw transmission data corresponding to the normalized transmission ratio shown in Figs.~3b and 3c of the main text.

While absolute optical transmission is not directly measured, the ratio of the optical transmission coefficients for forward and backward propagation can be obtained from the normalized microwave signal transmission coefficient $\bar T_\mu$, 
\be
|\TR/\TL|^2=|\bar T_{\mu R}/\bar T_{\mu L}|^2,
\ee
where $|\bar T_\mu|^2$ is normalized using the value of $|T_\mu|^2$ away from all mechanical resonances to remove all the external asymmetry in the experimental setup for left and right propagation paths.  These external asymmetries include modulator efficiency, cable/fiber loss, etc. In our analysis the normalization level is the average value of $|T_\mu|^2$ in the frequency range of $5.74$-$5.76$~GHz.  To be clear, the reason this calibration is necessary is because we don't actually physically swap the source and detector in our measurements.  Rather, for the left-to-right transmission path we have one modulator on the left side which generates the probe tone and one detector on the right side which measures the transmission through to the right side.  When we measure right-to-left transmission we have a different modulator on the right side to generate the probe tone and a different detector on the left side to detect the transmitted probe.  If the modulator on the left side is different from the modulator on the right side, then for the same microwave drive that excites the modulators we would get different a different optical probe power in the sidebands of the pump.  Similarly if the left and right detectors have different efficiencies then they would produce a different photocurrent for the same transmitted optical probe power.  Since we measure in practice the ratio of the microwave drive to the detected microwave photocurrent, this could cause an inherent asymmetry in the measured transmission for left-to-right and right-to-left transmission even if the \emph{optical} transmission was perfectly symmetric.

\subsection{Device characterization}

\begin{figure}[htp!]
\begin{center}
\includegraphics[width=0.7\columnwidth]{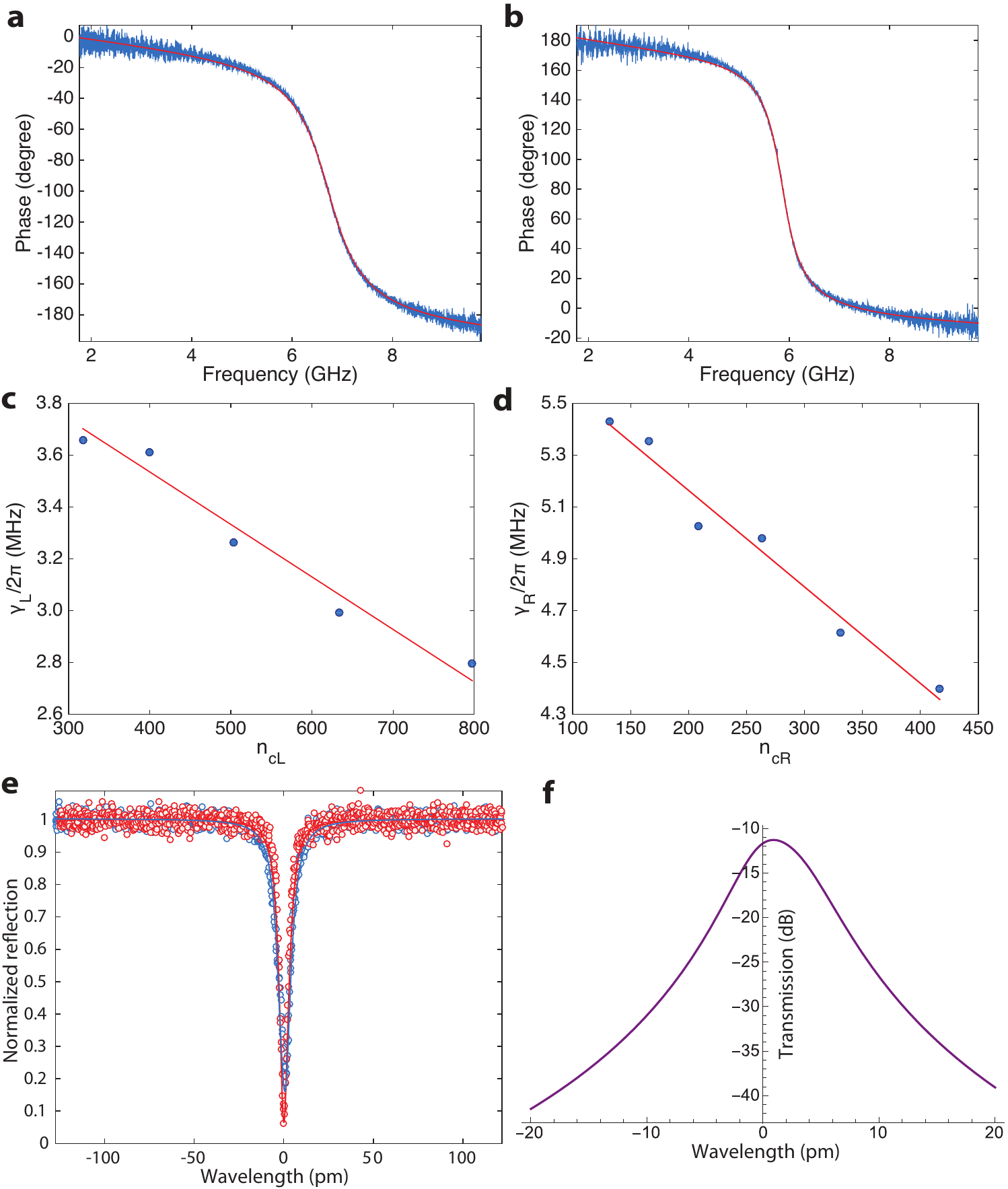}
\caption{  \textbf{a}, Left optical cavity phase response as measured by scanning the probe signal across the cavity resonance with weak blue-detuned pump.  \textbf{b}, Same as in \textbf{a} for the right optical cavity. \textbf{c}, Measured back-action modified mechanical linewidth versus intra-cavity pump photon number for the left optical cavity.  Here only left cavity pump beam is applied, and the pump is tuned to the upper motional sideband of the cavity (blue-detuned with $\Delta = +\omegamL$).  \textbf{d},  Same as in \textbf{c} for the right-side cavity and right-side pump. Measurements in \textbf{a}-\textbf{d} were performed prior to nano-oxidation tuning.  \textbf{e} Measured (circles) and theoretical (solid curves) optical reflection spectra using a left-cavity (blue) and right-cavity (red) optical pump.  These measurements are taken after nano-oxidation and the theoretical calculation includes the fit coupling ($J/2\pi=110$~MHz) between the left and right optical cavity modes and a splitting between the uncoupled modes. The wavelength origin is taken to correspond to the right optical cavity resonance.  \textbf{f} Calculated optical transmission power from one optical port to the other of an optical probe signal near resonance of the coupled optical cavity modes.  Here there is no pump beam, and so no coupling to phonons.  The parameters of the optical cavity modes are taken from the fit to the measured optical reflection spectra in \textbf{e}.}
\label{figs2}
\end{center}
\end{figure}

To determine the components of optical cavity loss (intrinsic decay rate $\kappai$, external waveguide-to-cavity coupling $\kappae$, total cavity decay rate $\kappa$) of both the left and right optical cavities we used a pump-probe scheme similar to that used to measure the nonreciprocity of the optomechanical circuit.  The pump beam in this case, however, is set to be very weak so as to not resonantly excite the mechanics as the probe signal is swept across the optical cavity resonance.  The cavity scans are plotted in Fig.~\ref{figs2}a and \ref{figs2}b for the left and right cavities, respectively. We fit the phase response curves and get $\kappaiLR/2\pi=0.29$ $(0.31)$~GHz, $\kappaeLR/2\pi=0.74$ $(0.44)$~GHz, and $\kappaLR/2\pi=1.03$ $(0.75)$~GHz. The intrinsic and external optical cavity rates are used to determine the intra-cavity photon number for a given optical pump power (specified at the input to the cavity).

Thermal mechanical spectra of the two cavities are measured with a weak blue-detuned optical pump so as to avoid back-action; a single pump is used for each of the left and right cavity measurements. The reflected pump light from the cavity contains modulation sidebands from the thermal mechanical motion, which upon detection with a high-speed photodetector creates a photocurrent with the thermal motion of the mechanical cavity modes imprinted on it. Since the mechanical modes can be hybridized between left-cavity, right-cavity, and waveguide modes, a measurement with the left-side pump produces a local measurement of the cavity modes as measured by the localized left optical cavity mode, and similarly for the right-side pump and cavity.  The intrinsic decay rate of the mechanical modes is inferred from the linewidth of the Lorentzian mechanical spectrum.

Measurements of the mechanical mode spectra were performed both before and after the cavities were nano-oxidized to tune their localized optical and mechanical modes into resonance.  Measurements prior to nano-oxidation allowed us to determine the local (left and right) mechanical and optical cavity mode properties (i.e., the bare, uncoupled mode properties).  Knowing the left and right cavity mode properties from independent measurements allowed us to fit with fewer fitting parameters the measured forward and backward transmission curves of the hybridized cavities presented in the main article text.  Note that after nano-oxidation the left and right optical cavity modes were only very weakly hybridized so as to maintain their left-cavity and right-cavity character.  The mechanical modes were tuned to be strongly hybdridized as evidenced in Fig.~2f of the main text.  Figures~\ref{figs2}c and \ref{figs2}d show the measured linewidth of the mechanical cavity modes $\MLR$ versus optical pumping power.  In Fig.~\ref{figs2}c the left cavity was pumped with a blue detuning $\Delta = +\omegamL$; in Fig.~\ref{figs2}d the right cavigty was pumped with a blue detuning of $\Delta = +\omegamR$.  By fitting the measured data with formula $\gamma=\gammai-4g_0^2n_c/\kappa$ ($n_c$ corresponding to the intra-cavity photon number determined from the $\OLR$ measured cavity properties), we obtain  $\gzeroLR/2\pi=0.76$ $(0.84)$~MHz and $\gammaiLR/2\pi=4.3$ $(5.9)$MHz for the left (right) localized cavity modes.

The optical ($J$) and mechanical ($V$) hopping rates between the two optomechanical cavities via the connecting waveguide are determined from a global fitting using Eq.~(1) of the main text for the group of measured transmission coefficient ratio curves in Figs.~3c and 3d with varying $\PhiB$.  The intra-cavity cavity photon number, optomechanical coupling rates and intrinsic mechanical decay rates are all taken as fixed and equal to the independently measured values as described above. 

With the fit value of $J$ from forward and reverse transmission measurements versus $\PhiB$, and the measured cavity coupling rates ($\kappa$, $\kappai$) from the left and rigth optical cavity modes prior to nano-oxidation tuning, we fit the measured optical reflection spectra of the two weakly coupled optical cavity modes after nano-oxidation.  This allows us to determine the uncoupled left and right optical cavity mode frequencies.  The measured and fit spectra as measured from the left and right cavities are shown in Fig.~\ref{figs2}e. As noted earlier, the measured spectra after nano-oxidation are still predominantly given by uncoupled left and right cavity modes.  Based on the theoretical fit to the measured optical reflection spectra, we also calculate the transmission of an optical probe signal through the optomechanical circuit in the absence of a pump beam (i.e., no coupling to phonons, just pure optical transmission)
\be\label{insertion}
\eta=\frac{J\sqrt{\kappaeL\kappaeR}}{|J^2+\kappaL\kappaR/4-(\omega-\omegacL)(\omega-\omegacR)-i\kappaL(\omega-\omegacL)/2-i\kappaR(\omega-\omegacR)/2|}.
\ee
Fig.~\ref{figs2}f shows the numerical result, and the minimum insertion loss for transmission from one port to the other port is found to be about $11$~dB for a probe signal frequency in between the two cavity resonances.  This is the estimated port-to-port optical transmission effiency in absence of optomechanical amplification.  


\section{Theory of optical nonreciprocity}
\label{App:B}

\subsection{Input-output formula}
We provide theoretical analysis of optical nonreciprocity in the coupled optomechanical cavity system. We first consider the case with two optical and two mechanical cavity modes. The Hamiltonian of this system can thus be written down as follows,
\begin{align} \label{H2}
\hat H     = & \sum_{k=L,R} \hbar \omega_{ck} \hat a_k^\dagger \hat a_k + J (\hat a_L^\dagger \hat a_R + \hat a_L \hat a_R^\dagger)
              +\sum_{k=L,R} \hbar \omega_{mk} \hat b_k^\dagger \hat b_k + V (\hat b_L^\dagger \hat b_R + \hat b_L \hat b_R^\dagger)
\\ \nonumber &
              +\sum_{k=L,R}  \hbar g_{0k} (\hat b_k^\dagger+\hat b_k) \hat a_k^\dagger \hat a_k 
              +\sum_{k=L,R} i\hbar \sqrt{\kappa_{ek}}\alpha_{pk}e^{-i\omega_{p k} t-i \phi_k}(\hat a_k-\hat a_k^\dagger), 
\end{align}
where $J$ and $V$ are the waveguide mediated optical and mechanical coupling strength (we gauged out the phase of $J$ and $V$ and take both of them to be real), and the last two terms are the optical driving fields (pumps) which have the same frequency and correlated phases.
%
We consider the situation where the optical cavities are nearly degenerate, i.e., $\omega_{cL} \simeq \omega_{cR} \equiv \omegac $ and both optomechanical systems are 
driven with a blue-detuned laser ($\omega_{p k} = \omegac+\omega_{mk} $). We perform a displacement transformation $\hat a_k = \alpha_{k } + \hat d_{k}$, separating the classical
steady state amplitude of the local optical cavity field from its fluctuations.  
With this we can linearize the optomechanical interaction in the Hamiltonian of Eq.~\ref{H2} in the usual manner. 
Assuming the good cavity limit (sideband resolved, $\omega_{mk} \gg \kappa_{k}$), we apply a rotating wave approximation and obtain for the equations of motions ($\hbar =1$)
\begin{align} \label{Eq.FullEoM}
 \frac{d}{dt} \hat d_L =& \left( i \Delta_{L}  - \frac{\kappa_{L}}{2} \right) \hat d_{L} 
                          - \sqrt{\kappa_{eL}}  \hat d_{L,\rm in}   - \sqrt{\kappa_{iL}} \hat \xi_{L,\rm in } 
                          - i J \hat d_{R} - i  G_{L}  \hat b_{L}^{\dag}   e^{i\phi_L},
                         \nonumber \\
 \frac{d}{dt} \hat d_R =& \left( i \Delta_{R}   - \frac{\kappa_{R}}{2}\right) \hat d_{R} 
                         - \sqrt{\kappa_{eR}} \hat d_{R,\rm in }   - \sqrt{\kappa_{iR}} \hat \xi_{R,\rm in } 
                         - i J  \hat d_{L} - i  G_{R}  \hat b_{R}^{\dag}   e^{i\phi_R},
                         \nonumber \\
 \frac{d}{dt} \hat b_L =& -\left(i \omegamL  + \frac{\gammaiL}{2}\right) \hat b_{L}   
                         - \sqrt{\gammaiL} \hat b_{L,\rm in }  
                         - i V \hat b_{R} 
                         - i  G_{L}  \hat d_{L}^{\dag}      e^{ i\phi_L},
                           \nonumber \\
 \frac{d}{dt} \hat b_R =& -\left( i \omegamR  + \frac{\gammaiR}{2} \right)\hat b_{R} 
                         - \sqrt{\gammaiR} \hat b_{R, \rm in }  
                         - i V \hat b_{L} 
                         - i  G_{R}  \hat d_{R}^{\dag}      e^{ i\phi_R},
\end{align}
with the total damping rates $\kappa_k = \kappa_{ek}+\kappa_{ik}$, the detunings $\Delta_k=\omegap-\omega_{ck}$ and the many-photon optomechanical couplings $G_{k} =  g_{0k}\alpha_k $.
The latter contains the steady state amplitude of the local optical cavity field $\alpha_ke^{i\phi_k}$, 
which is related to the pump amplitudes through 
\begin{align}
\alpha_{L(R)}e^{i\phi_{L(R)}} =& \frac{(i\Delta_{R(L)}-\kappa_{R(L)}/2)\sqrt{\kappa_{eL(R)}}\alpha_{pL(R)}e^{-i\varphi_{L(R)}}+iJ\sqrt{\kappa_{eR(L)}}\alpha_{pR(L)}e^{-i\varphi_{R(L)}}}
                                      {(i\Delta_L-\kappaL/2)(i\Delta_R-\kappaR/2)+J^2}.
\end{align}
\noindent We find the steady state amplitude is approximately $\sqrt{\kappa_{ek}}\alpha_{pk}e^{-i\varphi_{k}}/i\Delta_k$ under the condition $\Delta_k\approx\omega_{mk}\gg \kappa_k, J$, which means each cavity is effectively only driven by its own optical pump. 
Thus, each pump-enhanced optomechanical coupling and its phase can be independently controlled. 
The intrinsic noise operators $\hat \xi_{k,\rm in }$ and $\hat b_{k,\rm in }$ in the coupled mode equations \ref{Eq.FullEoM} describe thermal and vacuum fluctuations impinging on the the cavities and the mechanical modes respectively. The associated noise of a possible input signal is described via $\hat d_{k,\rm in}$. 

\subsection{Mechanically-mediated coupling}
We perform a Fourier transform ($\hat{b}[\omega] \equiv \int \text{d} t \, \hat{b}(t) e^{+i \omega t}$; $\hat{b}(t) \equiv \int \frac{\text{d} \omega}{2\pi} \, \hat{b}[\omega] e^{-i \omega t}$) of the coupled mode equations Eqs.~\ref{Eq.FullEoM} and insert the resulting solution for $\hat b^{\dag}_{L,R}[\omega]$  into the equations of the cavity operators. 
Ignoring the intrinsic noise terms $\hat \xi_{\rm in,k}$ and $\hat b_{\rm in,k}$ for the moment,
we obtain for the cavity operators in frequency space ($\PhiB = \phi_L -\phi_R$)
\begin{align} \label{Eq.:CavityOp}
\widetilde \chi_{L,+}^{-1}[\omega]  \hat d_{L}[\omega] =& - \sqrt{\kappa_{eL}} \hat d_{L,\rm in}[\omega] 
                                  - i \left( J -  \Gamma_{+}[\omega] e^{+ i \PhiB }   \right) \hat d_R[\omega]  ,
 \nonumber \\ 
 \widetilde \chi_{R,+}^{-1}[\omega] \hat d_{R}[\omega] =&  - \sqrt{\kappa_{eR}} \hat d_{R,\rm in}[\omega]
                                  - i  \left( J - \Gamma_{+}[\omega] e^{-i \PhiB}  \right) \hat d_{L}[\omega] ,
\end{align}
with the modified susceptibility $\widetilde \chi_{k,+}^{-1}[\omega] = \left( -i ( \omega + \Delta_k)   + \frac{\kappa_k}{2}   + i \Sigma_{k,+} [\omega]\right) $. The 
frequency dependent coupling $\Gamma_{+}[\omega]$ and the  self-energy $\Sigma_{k,+} [\omega]$  are defined as  
\begin{align}
 \Gamma_{+}[\omega] =        \frac{   V G_{R}  G_{L}      }{ \left[ - i (\omega + \omegamL) + \frac{\gammaiL}{2} \right]  \left[( - i ( \omega + \omegamR) + \frac{\gammaiR}{2} \right] + V^2}  ,
 \hspace{0.5cm}
  \Sigma_{k,+} [\omega] =&    \frac{ i G_{k} }{  V G_{\bar{k}}   }   \left[ - i ( \omega + \omega_{m\bar{k}})  +  \frac{\gamma_{i\bar{k}}}{2} \right]   \Gamma_{+}[\omega]  , 
\end{align} 
here the coupling $\Gamma_{+}[\omega]$ coincides with Eq.~(2) of the main text.
After eliminating the mechanical degrees of freedom, one finds both a "local" modification of each cavity (described by the self energy $\Sigma_{k,+}[\omega]$) and an induced coupling between the cavities.
The self-energies lead to damping (or anti-damping) of each cavity resonance as well as a frequency shift of the resonance. 
Here the subscript $+$ indicates blue-detuning ($\Delta_k =\omega_{pk} - \omegac \approx +\omega_{mk}$). The poles of the self energy read 
\begin{align}
 \omega_{\pm} = - \frac{i}{4} \left( \gammaiL + \gammaiR \right) - \frac{1}{2} \left( \omegamL +  \omegamR \right) \pm 
                \sqrt{   V^2 - \left[ \frac{1}{4} \left(\gammaiL - \gammaiR \right)   -   \frac{i}{2} (\omegamL - \omegamR)    \right]^2  }  .
\end{align} 

The induced coupling  has a coherent and a dissipative aspect. 
To illustrate this we separate the coupling into real and imaginary parts $\Gamma_{+}[\omega]  \equiv \Gamma_{\Re}[\omega]  + i \Gamma_{\Im}[\omega]$.  
The real and imaginary parts of this frequency-dependent coupling have completely different physical interpretations.   
We see this, by considering again the coupling terms in Eq.~(\ref{Eq.:CavityOp}). We have
\begin{align} \label{Eq.EffCoupling}
  \hat d_{L}[\omega]  \sim &   \ \left[ -i \left( J  - \Gamma_{\Re}[\omega] e^{+  i  \PhiB }   \right)   -    \Gamma_{\Im}[\omega] e^{+ i \PhiB } \right]  \hat d_R[\omega]  
               \equiv  \ \left[ -i \widetilde J^{\phantom{\ast}}[\omega]  -    \Gamma_{\Im}[\omega] e^{+ i \PhiB}  \right]  \hat d_R[\omega]   ,
 \nonumber \\ 
 \hat d_{R}[\omega]   \sim&    \ \left[ -i  \left(J   - \Gamma_{\Re}[\omega]  e^{-i  \PhiB }    \right)    -\Gamma_{\Im}[\omega] e^{-i \PhiB }  \right] \hat d_{L}[\omega]  
                \equiv \ \left[-i   \widetilde J^{\ast}[\omega]    -   \Gamma_{\Im}[\omega] e^{-i \PhiB }  \right] \hat d_{L}[\omega]  .
\end{align} 
For the given frequency of interest, we see that the real part of the induced coupling is completely equivalent to having a Hamiltonian, coherent tunneling term between the cavities;
we can absorb it into a redefinition of the coherent hopping strength $J$, i.e., $J \rightarrow \widetilde J[\omega]$. 
In contrast,  the coupling mediated by the imaginary part $\Gamma_{\Im}[\omega]$ \textit{is not} equivalent to some effective coherent tunneling interaction between the cavities,
i.e., the $\Gamma_{\Im}[\omega]$  terms in $\hat d_L$ and $\hat d_{R}$  Eqs.(\ref{Eq.EffCoupling}) cannot be incorporated into a definition of $J$.
The terms involving $\Gamma_{\Im}[\omega]$ instead represent a \textit{dissipative} coupling between the two cavities mediated by the mechanics.
Such dissipative interactions (if we ignore their frequency dependence) can be obtained in a master equation formalism via an effective Lindblad dissipator
of the form $2 \Gamma_{\Im}  \mathcal L \left[d_L^{\dag} + e^{-i \Delta \phi } d_R^{\dag} \right] $,
where $\mathcal L [\hat o]\hat \rho = \hat o \hat \rho \hat o^{\dag} - 1/2 \hat o^{\dag} \hat o \hat \rho - 1/2 \hat \rho \hat o^{\dag} \hat o  $ is the standard Lindblad superoperator.

\subsection{Directionality by balancing coherent and dissipative interactions}
The dissipative coupling is crucial for directionality: by balancing the dissipative interaction against the coherent interaction we obtain a nonreciprocal system
 (following the general recipe outlined in Ref.\cite{reservoir}).
For example, if we aim for a directional transport from the left to the right cavity, we want to decouple the left cavity from the right cavity (while still having the right cavity influenced 
by the left cavity). This is accomplished by balancing  coherent and dissipative interactions, i.e.,
\begin{align}\label{Eq.DirCond1}
 \widetilde J[\omega]    \overset{!}{=}  i  \Gamma_{\Im}[\omega] e^{i \PhiB}  ,
\end{align}
in which case the coupling from the left to right cavity vanishes,  cf. Eq.~(\ref{Eq.EffCoupling}), and we obtain a unidirectional coupling where the right cavity is driven
by the left cavity but not vice versa. Crucially, this would not be possible without the dissipative interaction, i.e., we need $\Gamma_{\Im}[\omega] \neq 0$.
Note, for the situation that $\Gamma_{\Im}[\omega] = 0$, i.e., $\gamma_{ik} = 0$, but finite $\PhiB$, we  still obtain a directional
dependent phase. However, to use this as the basic for nonreciprocal transmission additional interference processes have to implemented.

The directionality condition Eq.~(\ref{Eq.DirCond1}) can be reformulated in terms of the original $J$ and the phase difference $\PhiB$ as used in Eq.~(\ref{Eq.:CavityOp}).
This translates to the condition 
\begin{align}\label{Eq.DirCond2}
 J = \left|\Gamma_{+}[\omega] \right| ,
 \hspace{0.5cm}
   \PhiB =     - \arg(\Gamma_{+}[\omega] ) ,
\end{align}
where we still aim for unidirectional behavior from left to right.
For the case of a purely real coupling $\Gamma_{+}[\omega]= \Gamma_{\Re}[\omega]$ these conditions could still be satisfied, i.e., for $\PhiB = 0$ and $\Gamma_{\Re}[\omega] = J$. 
However, this means that there is effectively no coupling between the cavities, and thus no forward transport either. 
Note, that a sign change in $\arg(\Gamma_{+}[\omega] )$ would lead to the opposite situation, where the propagation direction would be from right to left.

In general, the directionality balancing condition obtained here is frequency dependent, for the simple reason that the mechanically-mediated cavity-cavity coupling is frequency-dependent.
If we could somehow fulfill the directionality condition in Eq.~(\ref{Eq.DirCond2}) at every frequency,
the cavity output field operators would be given by (using the standard input-output relation $\hat d_{k, \rm out} = \hat d_{k, \rm out} + \sqrt{\kappa_{ek}} \hat d_{k}$)
\begin{align} \label{Eq.:CavityOpDir}
   \hat d_{L,\rm out}[\omega] =& \left[1  -  \kappa_{eL}  \widetilde \chi_{L,+}[\omega] \right]\hat d_{L,\rm in}[\omega]   ,
 \nonumber \\ 
  \hat d_{R,\rm out}[\omega] =&  \left[1 -  \kappa_{eR}  \widetilde\chi_{R,+} [\omega] \right] \hat d_{R,\rm in}[\omega] 
                                  - i \sqrt{\kappa_{eR}\kappa_{eL}} \widetilde\chi_{R,+} [\omega] \widetilde\chi_{L,+}[\omega]\left|\Gamma_{+}[\omega] \right| 
                                       \left( e^{i  2\arg(\Gamma_{+}[\omega])     } -  1 \right)  
                                         \hat d_{L,\rm in}[\omega]   ,
\end{align} 
where we neglected the noise  contributions originating from the mechanical modes, i.e., the coupling to $\hat b_{n,\rm in}$ in Eq.~(\ref{Eq.:CavityOp}), 
and the intrinsic cavity noise $\xi_{\rm in,k}$ for simplicity.
Here, we see again that the dissipative interaction is crucial as we need $\arg(\Gamma_{+}[\omega]) \neq n \pi, n \in \mathbb{Z}$,  i.e., we need a finite imaginary part of $\Gamma_{+}[\omega]$.

The  experimentally relevant situation is where dissipative and coherent interactions are only balanced at a single frequency 
(by appropriate tuning of phase and $J$). Achieving this condition close to the normal modes resonance frequencies is favorable given the resonantly-enhanced transmission.
Enforcing directionality at $\omega = - \omegam  \pm  V $ for equal mechanical resonance frequencies, results in the directionality conditions
 \begin{align}\label{Eq.DirCond3}
 \omegamL=\omegamR :
 \hspace{0.5cm}
 \PhiB = \mp \arctan   \frac{2 V \left(\gammaiL +\gammaiR \right)}{ \gammaiL\gammaiR } ,
 \hspace{0.5cm}
 J =   \frac{   V G_{R}  G_{L}  }{\sqrt{ \frac{1}{4}  V^2    \left(\gammaiL + \gammaiR \right)^2 +   \frac{\gammaiL^2\gammaiR^2}{16} } }   , 
\end{align}
where the upper (lower) sign in the phase difference $\PhiB $ realizes directionality at $\omega = -\omegam + V  (-\omegam - V )$. 
Directionality here means  that an input signal  injected on the left cavity is transmitted to the right cavity, whereas the 
backward propagation path, i.e., from right to left, is blocked.  

On the other side, if we assume identical bare mechanical damping of the mechanical modes ($\gammaiL=\gammaiR=\gammai$), but unequal bare mechanical frequencies ($\omegamL \neq \omegamR$), 
then we find that at the frequencies of the hybridized mechanical modes  $\Omega_{\pm} =    - \frac{1}{2} \left( \omegamL +  \omegamR \right) \pm   \sqrt{   V^2 +   \frac{1}{4} (\omegamL - \omegamR)^2  }$
the directionality condition is modified to
 \begin{align}\label{Eq.DirCond4}
 \gammaiL=\gammaiR :
 \hspace{0.5cm}
 \PhiB = \mp \arctan   \frac{4 \sqrt{  V^2+\frac{1}{4}(\omegamL-\omegamR)^2} }{ \gamma}  ,
 \hspace{0.5cm}
 J =  \frac{     V G_L G_R }{\gamma  \sqrt{  V^2 + \frac{\gamma^2}{16} + \frac{1}{4}(\omegamL-\omegamR)^2 } } . 
\end{align}
where the upper (lower) sign in the phase difference $\PhiB $ realizes directionality at $\omega = \Omega_{+ (-)}$. 
The directionality conditions for a perfectly symmetric device, i.e., for equal mechanical resonance frequencies $(\omegam)$ and decay rates $(\gamma)$, can simply be read off from
either Eq.~\ref{Eq.DirCond4} or Eq.~\ref{Eq.DirCond3}. 

\subsection{Nonreciprocal optical transmission: two blue-detuned pumps}
From the equations for the cavity operators in Eqs.~\ref{Eq.:CavityOp} we can calulate the transmission coefficients via input/output theory.
Note, that although Eqs.~\ref{Eq.:CavityOp} are formulated on the basis of noise operators, they as well describe the dynamics of the cavity field amplitudes $d_{k}$
around their steady state solution.
The right transmission coefficient $\TR \equiv d_{R,\rm out}/d_{L,\rm in}$ and left transmission coefficient $\TL\equiv d_{L,\rm out}/d_{R,\rm in}$ are given by
\begin{align}
T_{R \leftrightarrows  L} [\omega] =& \frac{i \sqrt{\kappaeL \kappaeR}    \left[ J - \Gamma_{+}[\omega]  e^{\mp i \PhiB }\right]}
                                    { \widetilde \chi_{L}^{-1}[\omega]  \widetilde \chi_{R}^{-1}[\omega] +  \left[ \Gamma_{+}[\omega]^2 + J^2 - 2 \Gamma_{+}[\omega] J \cos (\phi )\right]   }
                                    \equiv A_{+}[\omega]  \left[ J - \Gamma_{+}[\omega]  e^{\mp i \PhiB }\right],  
\end{align}
with the modified susceptibilities $\widetilde \chi_{k}[\omega]$ as defined after Eq.~(\ref{Eq.:CavityOp}). 
The prefactor $ A_{+} [\omega]$ is the same for both transmission amplitudes, it accounts for the mechanically-induced back-action on the optical cavities, cf. main text after Eq.~(2).  
Note, that the corresponding prefactor for two red-detuned pumps is simply  $A_{-}[\omega] = - A_{+}^{\ast}[-\omega]$.

We now assume a completely symmetric pair of mechanical cavities ($\omegamL=\omegamR=\omegam$ and $\gammaiL=\gammaiR=\gammai$) and apply the 
corresponding directionality direction for symmetric parameters, cf. Eq.~\ref{Eq.DirCond4} or Eq.~\ref{Eq.DirCond3}.
The transmission coefficient for the through direction ($\rightarrow$) under these conditions of perfect nonreciprocity is given by,
\begin{align}\label{FullSolTrans}   
 T_{\rightarrow} [-\omegam \pm V]     =& \sqrt{\frac{\kappaeL \kappaeR}{\kappa _R \kappa _L}}
       \sqrt{ \frac{1 \pm i\frac{\gammai}{ 4V}      }{1 \mp i \frac{\gammai}{ 4V}    }}
       \frac{   8  i  \sqrt{  \mathcal C_L  \mathcal C_R  }  }
           {     
           \left[    \mathcal C_L \left(1 \pm i \frac{\gamma_i}{2V}  \right)    -  \left(1 \mp i \frac{2 V}{\kappa _L}\right)    \left(2 \pm i \frac{\gamma_i}{2V}  \right) \right] 
           \left[    \mathcal C_R \left(1 \pm i \frac{\gamma_i}{2V}  \right)    -  \left(1 \mp i \frac{2 V}{\kappa _R}\right)    \left(2 \pm i \frac{\gamma_i}{2V}  \right) \right]},
\end{align}
introducing the single cavity cooperativity $C_k \equiv 4G_k^2/\gammai\kappa_k$. 
Considering as well symmetric optical cavities ($\kappaeL=\kappaeR=\kappae$; $\kappaL=\kappaR=\kappa$) with symmetric optical pumping ($\GL=\GR=G$)
the transmission coefficient simplifies to
\begin{align}\label{maxtransblue} 
T_{\rightarrow} [- \omegam \pm V]  &   \overset{V \ll \kappa  }{\simeq}  
                                          \frac{8 i \mathcal C  \frac{\kappa_{e}}{\kappa}  }
                                               { \left[  2  -   \mathcal C \pm i   \frac{\gammai}{2V} \left( 1 - \mathcal C  \right)      \right]^2 },  
\end{align}
with $C  \equiv 4G^2/\gammai\kappa$ and under the realistic assumption that the hopping rate $V$ is much lower than the cavity decay rate $\kappa$.
Here we work with blue-detuned pumping of both optical cavities ($\Delta \approx +\omegam$), which results in parametric amplification of each of the left and right mechanical modes  
and leads to amplification of the optical probe signal.
This becomes apparent for the situation that the mechanical hopping rate is much faster than the intrinsic mechanical decay rate ($V/\gammai\gg 1$). 
In this case the gain diverges for $\mathcal C \rightarrow 2$ (this is twice as large as for a single cavity instability because the mechanical modes are hybridized and thus the effective optomechanical coupling from the left or right optical cavity is reduced by a factor of $\sqrt{2}$, hence the cooperativity by a factor of 2).  Note, for the situation $V/\gammai\gg 1$,
 the directionality conditions at the hybridized mechanical modes $\omega= - \omegam \pm V$ simplifies to $J \simeq  \GL\GR/\gammai$ and $\PhiB \rightarrow \mp \pi/2$.

\subsection{Nonreciprocal optical transmission: two red-detuned pumps}
The analysis for the the case of two red detuned pumps is similar to the blue-detuned case. 
The cavity operators in  Eq.(\ref{Eq.FullEoM}) couple now to the mechanical lowering operators $\hat b_{k}$ and vice versa, while
the detuning between the cavity resonances and the external pump tones yields $\Delta_k = - \omega_{mk}$. 
The ratio of transmission coefficients is found to be given by the following expression
\begin{align}
\frac{\TR}{\TL} =& \frac{J - \Gamma_{-}[\omega]e^{-i\PhiB} }{J - \Gamma_{-}[\omega]e^{+i\PhiB}} = 
\frac{J  - \frac{V  G_L G_R}{\left[ -i \left(\omega - \omegamL \right) + \frac{\gammaiL}{2}\right] \left[ - i\left(\omega - \omegamR \right)+\frac{\gammaiR}{2}\right] + V^2} e^{-i\PhiB} }
     {J  - \frac{V  G_L G_R}{\left[ -i \left(\omega - \omegamL \right) + \frac{\gammaiL}{2}\right] \left[ - i\left(\omega - \omegamR \right)+\frac{\gammaiR}{2}\right] + V^2} e^{+i\PhiB} }
\end{align}
where we have $\Gamma_{-}[\omega] = \Gamma_{+}^{\ast}[-\omega]$, thus the ratio $|\TR/\TL|$ is the same for blue and red detuned pumps evaluated at corresponding frequencies.
The reason for this is that the transmission is either amplified or suppressed simultaneously for both directions and thus their ratio stay unchanged.  
Comparing to the blue detuned case, the perfect nonreciprocity condition remains the same in the red detuned case, while the transmission coefficient for the through direction 
the hybridized mechanical modes  $\Omega_{\pm} =  \omegam \pm  V $ is given by (assuming $\omegamL=\omegamR$, $\gammaiL=\gammaiR=\gammai$ and $V\ll \kappa_k$)
\begin{align}  \label{maxtransred}
 T_{\rightarrow} [  \omegam \pm V] \simeq & \sqrt{\frac{\kappaeL \kappaeR}{\kappa _R \kappa _L}}  
 \frac{ 8 i  \sqrt{\mathcal C_L  \mathcal C_R   }}
      {   
       \left[    \mathcal C_L +    2  \pm i \frac{\gammai}{2V} \left( \mathcal C_L  +1 \right)     \right]
       \left[    \mathcal C_R +    2  \pm i \frac{\gammai}{2V} \left( \mathcal C_R  +1 \right)     \right]}.
\end{align}
From Eq.~\ref{maxtransred}, we note in general an attenuated transmission for the red detuned case as $T_{\rightarrow}\le\sqrt{\kappaeL\kappaeR/(\kappaL\kappaR)}<1$. For the case of a fast hopping rate $V/\gammai\gg 1$
 equality is achieved when $\mathcal C_{k} = 2$ and/or $\kappa_k/2=\frac{\GL\GR}{\gammai}$. Comparing the latter to Eq.~\ref{Eq.DirCond3} we see the maximal through transmission efficiency is achieved when the optical cavity loss rate $\kappa_k/2$ is matched to the inter-cavity photon hopping rate $J$ for both cavities (impedance matching condition).

\subsection{Nonreciprocity associated with a single mechanical waveguide mode}
In our optomechanical circuits, we also observed optical nonreciprocity with a single mechanical waveguide mode. In this case, the Hamiltonian describing the interaction between two optical cavity modes and one mechanical waveguide mode is given by,
\begin{align}\label{H3}
\hat H &  =    \sum_{k=\text{L,R}} \hbar\omegack \akdag\akhat  + J (\aLdag\aRhat + \aLhat\aRdag)+\hbar \omegamW\bWdag\bWhat
    \\ \nonumber &
             + \sum_{k=\text{L,R}} \hbar \left(\gzeroWk   \bWhat  + \gzeroWk^{\ast}   \bWdag     \right) \akdag\akhat
              +\sum_{k=\text{L,R}} i\hbar\sqrt{\kappa_{\text{e}k}}\alpha_{\text{p}k}e^{-i\omegap t-i\phi_k}(\akhat-\akdag).
\end{align}
Going through a similar calculation using coupled mode equations, we find that the ratio of right and left optical transmission coefficients is
\begin{align}\label{wratio}
\frac{\TR}{\TL} = &   \frac{J \pm i \frac{ |\GWL\GWR| }{-i\left(\omega \pm \omegamW\right) + \frac{\gammaiW}{2} }e^{- i\left(\PhiB \pm \Phi_{W} \right)}}
                           {J \pm i \frac{ |\GWL\GWR| }{-i\left(\omega \pm \omegamW\right) + \frac{\gammaiW}{2} }e^{+ i\left(\PhiB \pm \Phi_{W} \right)}},
\end{align}
where the upper (lower) sign corresponds to the blue (red) detuned case and $\Phi_{W} = \arg(\GWL^{\ast}\GWR )$. 
The corresponding conditions for perfect directionality from left to right and at $\omega = \mp \omegamW$   are
\begin{align}\label{singlecondition}
J =  \frac{2 |\GWL\GWR|}{  \gammaiW  }, 
\hspace{0.5cm}
 \PhiB = \pm \frac{\pi}{2} \mp \Phi_{W}. 
\end{align}
This in turn leads to the transmission coefficients
\begin{align}
  T_{\rightarrow} [ \mp \omegamW ] = &  \sqrt{ \frac{ \kappaeL \kappaeR }{\kappaL \kappaR } }  
                                  \frac{4 i  \sqrt{ \mathcal C_{WL} \mathcal C_{WR}  }  }
                                       {(\mathcal C_{WL} \mp 1) (\mathcal C_{WR} \mp 1)  } .
\end{align}
In the case of blue detuned tones an input signal is amplified and the corresponding gain increases for $\mathcal C_{Wk} \rightarrow 1$.

Note in Eq.~\ref{singlecondition} we included the phase of the product $\GWL^{\ast}\GWR$. This addition comes from the fact that we have already chosen definitions for the local cavity mode amplitudes ($a_{\text{L,R}}$ and $b_{\text{L,R}}$) such that the phase of the optomechanical couplings of the localized cavity modes -- $\GL\equiv |\alpha_{\text{L}}|\gzeroL$ and $\GR\equiv |\alpha_{\text{R}}|\gzeroR$ -- are both zero.  With these same definitions for amplitudes $a_{\text{L}}$ and $a_{\text{R}}$ we are not then free to set the phases of \emph{both} $\GWL$ and $\GWL$ to be zero; not at least for the same set of pump phases $\phi_{\text{L}}$ and $\phi_{\text{R}}$ chosen for the localized cavity mode coupling. A simple example helps to illustrate this.  The mode $\Mw$ can be viewed as a hybridization between the localized left and right cavity modes and a delocalized waveguide mode~\cite{omcc}.  Using perturbation theory, we have for the mechanical mode amplitude of the hybridized mode $\Mw$,
\be\label{hybridization}
\bW=\bWprime + \frac{\tL}{\omegamWprime - \omegamL}\bL+\frac{\tR}{\omegamWprime-\omegamR}\bR,
\ee 
where $\bWprime$ is the unperturbed delocalized waveguide mode amplitude and $\omegamWprime$ is the unperturbed frequency of the delocalized waveguide mode.  $t_{\text{L(R)}}$ is the coupling coefficient between the delocalized waveguide mode and the localized cavity mode $M_{\text{L(R)}}$. The phases of $\tL$ and $\tR$ are determined by the field distribution of the hybridized mode $\Mw$ in the left and right cavities, respectively, and cannot be (both) chosen arbitrarily . Using the mode decomposition of Eq.~\ref{hybridization}, we have that $\arg(\gzeroWL^{\ast}\gzeroWR) =  \arg(\tL^{\ast}\tR)$ as we have already chosen a local cavity mode amplitude basis such that $\arg(\gzeroL)=\arg(\gzeroR)=0$ and $\omegamWprime>\omegamL,\omegamR$ (this assumes of course that the left (right) optical cavity mode only couples to the portion of $\bW$ which is due to $\bL$ ($\bR$), which is a good approximation due to the fact that the optical cavities are in the far field of each other).  Thus, by \emph{simultaneously} measuring the flux-dependent transmission near the resonance of the localized mechanical cavity modes and the hybridized mechanical waveguide mode we can determine the $\arg(\gzeroWL^{\ast}\gzeroWR)$ in this mode basis (see Fig.~\ref{figs4} for example). For the $M_W$ mode in our experiment, we find $\arg(\gzeroWL^{\ast}\gzeroWR) \approx \pi$, which means for this hybridized mode and chosen localized cavity mode basis the mechanical motion in the left cavity as seen by the left cavity optical mode is approximately $180$ degrees out of phase with the motion in the right cavity as seen by the right cavity optical mode.


\section{Directional flow of quantum and thermal noise}
\label{App:C}
Besides the nonreciprocal optical signal transmission, the flow of quantum and thermal noise in the optomechanical circuit is directional. 
This is a natural consequence of the system's scattering matrix having a directional form; the scattering matrix determines both the transmission of coherent signals, as well as noise properties. 
To show this, we calculate the symmetrized output noise spectral density via
\begin{align}
 \bar S_{k,\rm out}[\omega] = \frac{1}{2} \int \frac{d\Omega}{2\pi} \ev{\left\{\hat d_{k, \rm out}[\omega], \hat d_{k, \rm out }^{\dag}[\Omega] \right\}},
\end{align}
defined in the standard manner \cite{Clerk2010}. The mechanical and optical noise  operators introduced in Eqs.~\ref{Eq.FullEoM} have zero mean
and satisfy the canonical correlation relations:
\begin{align}
\langle \hat o_{k, \rm in }[\omega] \hat o^\dagger_{k^\prime,\rm in} [\Omega] \rangle 
= \langle o^\dagger_{k, \rm in } [\omega] o_{k^\prime, \rm in }[\Omega] \rangle + \delta_{k,k^\prime}\delta(\omega+\Omega)
= \left( n_{o_k}^{\rm th}+1 \right) \delta_{k,k^\prime}\delta(\omega+\Omega),
\hspace{0.5cm} 
\hat o_{k, \rm in} = \hat d_{k, \rm in},\hat \xi_{k, \rm in},\hat b_{k, \rm in} .
\end{align}
where $n_{o_k}^{\rm th}$ is the thermal occupation of each bath. In what follows, we assume that we have no thermal occupation of the optical field.
This is justified as we work with a very high optical frequency. 

\begin{figure}[htp!]
\begin{center}
\includegraphics[width=0.7\columnwidth]{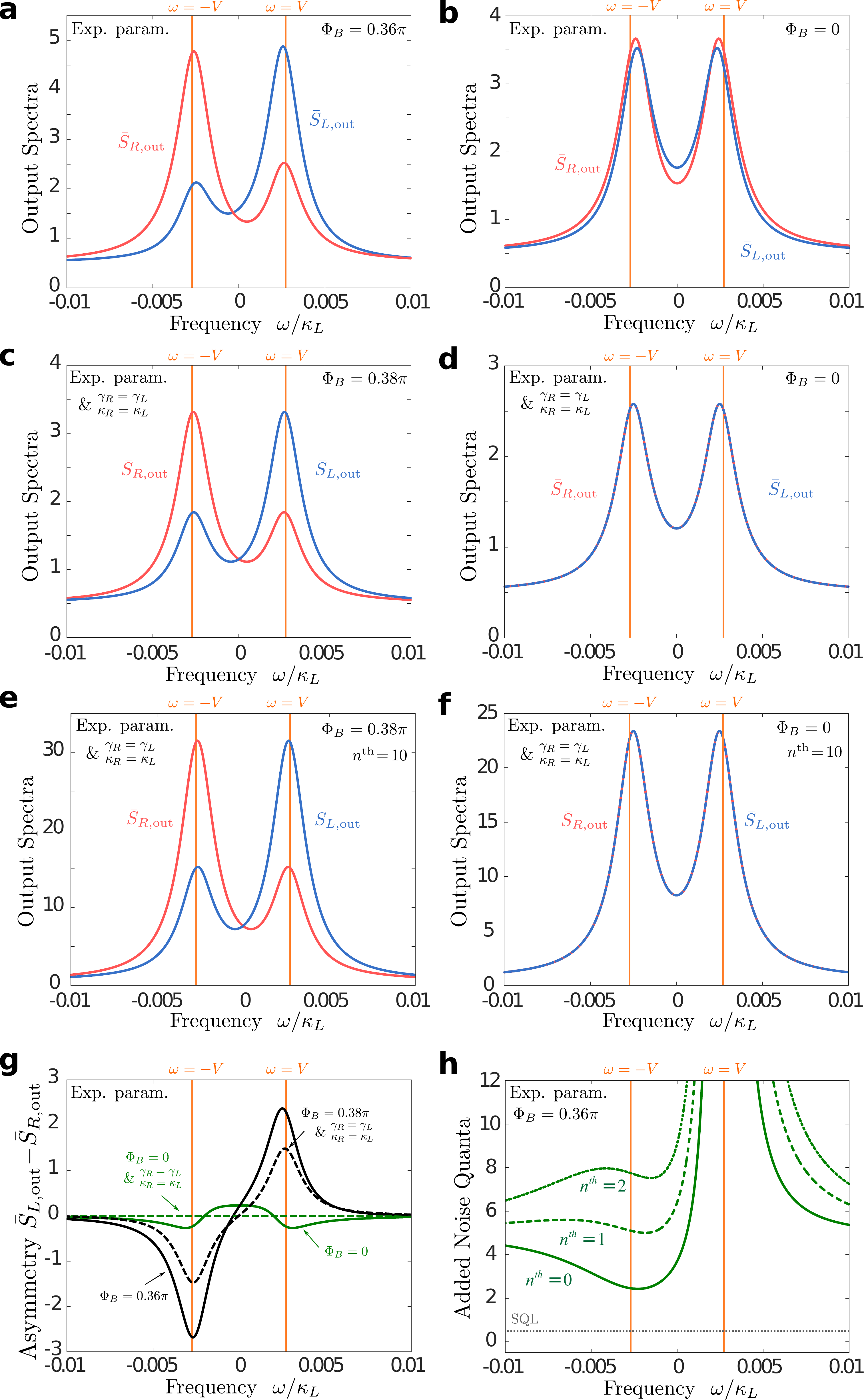}
\caption{ \textbf{a-d} Output noise spectra at zero temperature for a set of parameters given in the text. \textbf{e-f} Output noise spectra at finite temperature with thermal phonon occupation of $n^{\rm th}=10$. \textbf{g} Difference of the left and right output spectra for \textbf{a-d}. \textbf{f} Added noise for right-propagation signal. }
\label{figs3}
\end{center}
\end{figure}

Figure \ref{figs3}a-d depicts the output spectra for the situation that both pumps are blue detuned from the cavity by $\omegam$. 
Here we assumed equal mechanical frequencies $\omegamL=\omegamR=\omegam$ and  work in a rotating frame where the uncoupled mechanical resonance frequencies are shifted to zero.
The remaining parameters are as used in the experiment, i.e., we take $\gammaiL/2\pi=4.3$~MHz, $\gammaiR/2\pi = 5.9$~MHz, $\kappaL/2\pi = 1.03$~GHz, $\kappaR/2\pi = 0.75$~GHz, $\kappa_{iL}/2\pi = 0.29$~GHz, $\kappa_{iR}/2\pi = 0.31$~GHz, $V/2\pi = 2.8$~MHz, $J/2\pi = 110$~MHz. The multiphoton couplings $\GL=\GR$ used in the calculation are determined from Eq.~\ref{Eq.DirCond3}. 

Figure~\ref{figs3}a shows the result for zero temperature mechanical baths  and a finite phase $\PhiB=0.36\pi$ (determined from Eq.~\ref{Eq.DirCond3}). 
As expected, the $L$ and $R$ output spectra are not identical: while each has a double-peaked structure (corresponding to the two normal mode resonances), 
the right output spectra $\bar S_{R,\rm out}[\omega]$ has the upper-frequency peak larger than the lower-frequency peak, while the situation is reversed for
the left output spectra. This does not lead to any asymmetry in the total output photon number fluxes  (i.e., intergrated over all frequencies). It does however lead to an asymmetry in the energy fluxes 
(i.e., as the higher energy peak is bigger for the right output spectrum, and the low energy peak is bigger for the left spectrum).
Thus, the  "quantum heating" of zero-point fluctuations preferentially cause an energy flow to
the right (rather than to the left) for this choice of phase.   

It is also worth noting that if all dissipative rates are equal for the $R$ and $L$ cavities, then the $L$ output spectrum is just the frequency-mirrored $R$ output spectrum.
The latter is visible in Fig. \ref{figs3}(c), where we plotted the output spectra for symmetric parameters, i.e., we set $\gammaiR/2\pi=\gammaiL/2\pi=4.3$~MHz, $\kappaR/2\pi =\kappaL/2\pi = 1.03$~GHz, $\kappa_{iR}/2\pi =\kappa_{iL}/2\pi = 0.31$~GHz  and $\PhiB=0.38\pi$ (determined from Eq.~\ref{Eq.DirCond3} for the new $\gammaiR$).
However, having unequal decay rates, i.e., $\gamma_{R} \neq \gamma_{L}$ and $\kappa_{R} \neq \kappa_{L}$, leads to a slight asymmetry even if the phase is set to zero, i.e., $\PhiB = 0$, as visible in Fig.\ref{figs3}b.
In Fig.~\ref{figs3}g we plot the asymmetry $\bar S_{L,\rm out}[\omega]-\bar S_{R,\rm out}[\omega]$ for all the four cases corresponding to Fig.~\ref{figs3}a-d. 

For finite temperature, we find that the output spectrum has a roughly linear dependence on the mechanical bath temperature: $\bar S_{k, \rm out}(T)=c_k n^{\rm th}+\bar S_{k, \rm out}(0)$ 
(assuming $n_{b_L}^{\rm th}=n_{b_R}^{\rm th} \equiv n^{\rm th}$). This linear dependence is visible if we compare Fig.~\ref{figs3}c,d and Fig.~\ref{figs3}e,f,
where the latter  show the output noise spectra for $n^{\rm th}=10$ with symmetric cavity parameters.  
Additionally, we also calculate the added noise quanta to the transmitted signal
\begin{align}
\bar n_{k,\rm add}[\omega] \equiv \frac{\bar S_{k,\rm out}[\omega] }{|T_k [\omega]|^2}-\frac{1}{2},
\end{align}
where $\frac{1}{2}$ is the half quanta noise of the vacuum optical fields injected from the coupler. Fig.\ref{figs3}h shows the added noise for left-right propagation with $\PhiB=0.36\pi$ (and asymmetric experimental cavity parameters). The mechanical baths $n^{\rm th}$ are varied as denoted in each graph. Even if the cavities and the mechanics are only driven by vacuum noise the standard quantum limit (SQL) of half a quanta is not achieved.  This is due to the limited amount of gain achieved in the experiment, i.e., the transmission coefficient is not high enough to suppress the noise contributions. 
Moreover, even in the large gain limit the added noise would be roughly one quanta due to the finite amount of intrinsic optical cavity loss.

\section{Reciprocal device}
\label{App:D}

\begin{figure}[htp!]
\begin{center}
\includegraphics[width=0.7\columnwidth]{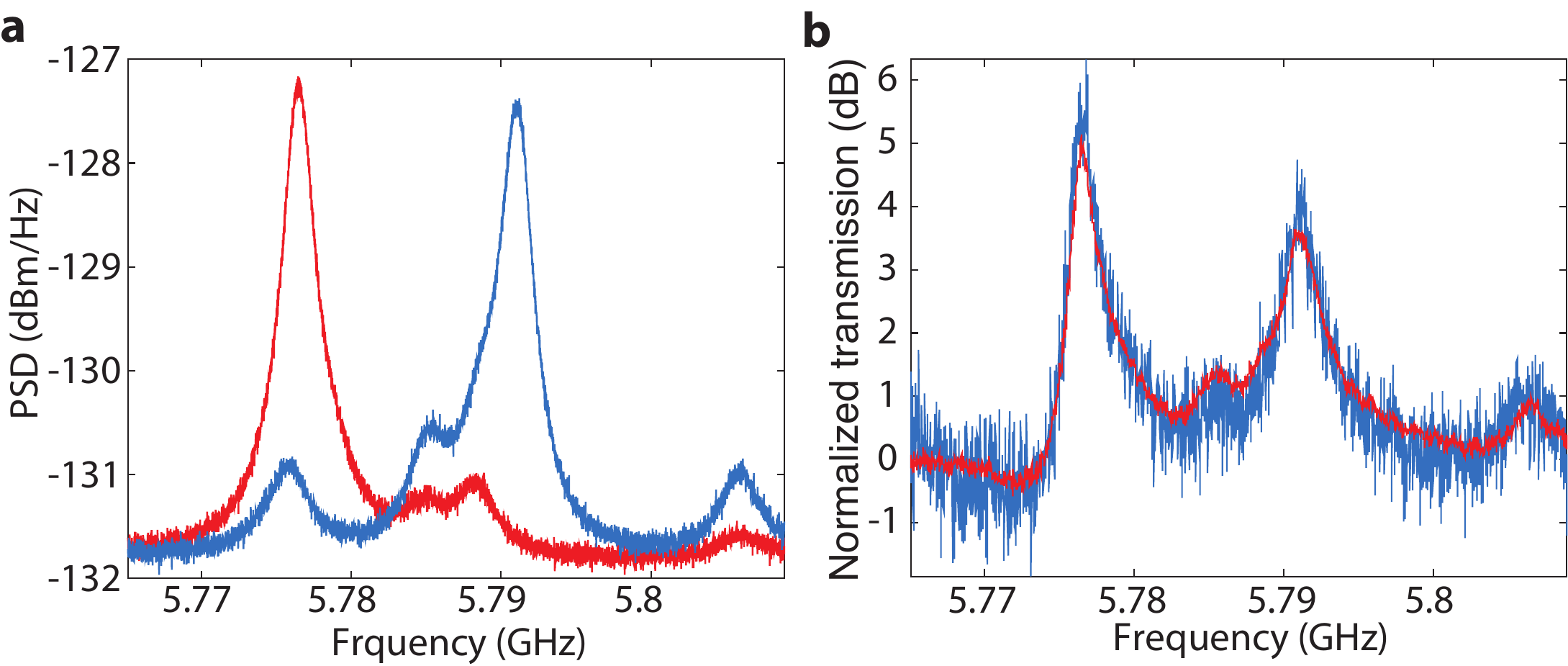}
\caption{ Optical reciprocity in a circuit with large optical cavity coupling, $J$. \textbf{a} Mechanical spectra measured from the left (red) and right (blue) optical cavities. \textbf{b} Normalized optical signal power transmission coefficient for forward (red) and reverse (blue) optical signal propagation. }
\label{figs4}
\end{center}
\end{figure}

Realizing optical nonreciprocity in the optomechanical circuits studied in this work is not simple or easy as just creating a circuit with optical and mechanical coupling between two optomechanical cavities.  One is limited by the practical realities of device power handling capability, finite optical and mechanical $Q$-factors, etc.  As such, not all the circuits that were tested exhibited nonreciprocal transmission and amplification; the effects were too weak to observe in some circuits.  This, however, was a useful test of our set-up as nonreciprocity could be effectively turned on and off by looking at different circuits with only slightly different parameters.

Eq.~\ref{Eq.DirCond4} sets the desired circuit parameters in order to achieve significant nonreciprocity, which for the optomechanical coupling, optical and mechanical $Q$-factors, and the power handling capabilities of the nanobeam cavities requires optical hopping rate between cavities to be less than $J/2\pi \approx 500$~MHz.  Devices with larger coupling rates can simply not be pumped hard enough to satisfy $\Gk \approx (J \gammaik)^{1/2}$. To confirm this, here we show another optomechanical crystal circuit with bare cavity wavelengths of $\lambda_{L(R)}=1535.051$ $(1535.060)$~nm and inter-cavity photon hopping rate of $J/2\pi=1.4$~GHz (more than ten times larger than the device studied in the main text). The mechanical spectra of this device as measured from both the left and right optical cavities is shown in Fig.~\ref{figs4}a.  Figure~\ref{figs4}b shows the normalized transmission coefficient for forward and reverse optical signal propagation for a blue-detuned pump wavelength of $\lambda_p=1534.99$~nm and synthetic flux of $\PhiB=\pi/2$.   Even at the largest pump powers ($P_{p} \approx 100$~$\mu$W; $\ncavO \approx 1.5\times 10^{3}$) this device does not satisfy the condition of Eq.~\ref{Eq.DirCond4} due to the large $J$, resulting in nearly perfect reciprocity in the optical signal transmitted power.  These measurements were performed on the exact same set-up as the circuit studied in the main text.   
 

\begin{thebibliography}{10}
\providecommand{\url}[1]{#1}
\csname url@samestyle\endcsname
\providecommand{\newblock}{\relax}
\providecommand{\bibinfo}[2]{#2}
\providecommand{\BIBentrySTDinterwordspacing}{\spaceskip=0pt\relax}
\providecommand{\BIBentryALTinterwordstretchfactor}{4}
\providecommand{\BIBentryALTinterwordspacing}{\spaceskip=\fontdimen2\font plus
\BIBentryALTinterwordstretchfactor\fontdimen3\font minus
  \fontdimen4\font\relax}
\providecommand{\BIBforeignlanguage}[2]{{%
\expandafter\ifx\csname l@#1\endcsname\relax
\typeout{** WARNING: IEEEtran.bst: No hyphenation pattern has been}%
\typeout{** loaded for the language `#1'. Using the pattern for}%
\typeout{** the default language instead.}%
\else
\language=\csname l@#1\endcsname
\fi
#2}}
\providecommand{\BIBdecl}{\relax}
\BIBdecl

\bibitem{manybody2}
M.~Schmidt, S.~Ke{\ss}ler, V.~Peano, O.~Painter, and F.~Marquardt,
  ``Optomechanical creation of magnetic fields for photons on a lattice,''
  \emph{Optica}, vol.~2, pp. 635--641, 2015.

\bibitem{Dalibard2011}
J.~Dalibard, F.~Gerbier, G.~Juzeliūnas, and P.~\"{O}hberg, ``Colloquium:
  Artificial gauge potentials for neutral atoms,'' \emph{Rev. Mod. Phys.},
  vol.~83, p. 1523, 2011.

\bibitem{Koch2010}
J.~Koch, A.~Houck, K.~L. Hur, and S.~Girvin, ``Time-reversal-symmetry breaking
  in circuit-qed-based photon lattices,'' \emph{Phys. Rev. A}, vol.~82, p.
  043811, 2010.

\bibitem{Devoret2015}
P.~Roushan, C.~Neill, A.~Megrant, Y.~Chen, R.~Babbush, etal, and J.~Martinis,
  ``Chiral groundstate currents of interacting photons in a synthetic magnetic
  field,'' p. arXiv:1606.00077, 2016.

\bibitem{Martinis2016}
K.~Sliwa, M.~Hatridge, A.~Narla, S.~Shankar, L.~Frunzio, R.~Schoelkopf, and
  M.~Devoret, ``Reconfigurable josephson circulator/directional amplifier,''
  \emph{Phys. Rev. X}, vol.~5, p. 041020, 2015.

\bibitem{Umucal2011}
R.~Umucal{\i}lar and I.~Carusotto, ``Artificial gauge field for photons in
  coupled cavity arrays,'' \emph{Phys. Rev. A}, vol.~84, p. 043804, 2011.

\bibitem{Hafezi2011}
M.~Hafezi, E.~Demler, M.~Lukin, and J.~Taylor, ``Robust optical delay lines
  with topological protection,'' \emph{Nat. Phys.}, vol.~7, p. 907, 2011.

\bibitem{fang}
K.~Fang, Z.~Yu, and S.~Fan, ``Realizing effective magnetic field for photons by
  controlling the phase of dynamic modulation,'' \emph{Nature Photon.}, vol.~6,
  pp. 782--787, 2012.

\bibitem{Rechtsman2013}
M.~Rechtsman, J.~Zeuner, Y.~Plotnik, Y.~Lumer, D.~Podolsky, F.~Dreisow,
  S.~Nolte, M.~Segev, and A.~Szameit, ``Photonic floquet topological
  insulators,'' \emph{Nature}, vol. 496, p. 196, 2013.

\bibitem{Tzuang2014}
L.~Tzuang, K.~Fang, P.~Nussenzveig, S.~Fan, and M.~Lipson, ``Nonreciprocal
  phase shift induced by an effective magnetic flux for light,'' \emph{Nature
  Photons}, vol.~8, p. 701, 2014.

\bibitem{Lin2009}
Y.-J. Lin, R.~Compton, K.~Jiménez-García, J.~Porto, and I.~Spielman,
  ``Synthetic magnetic fields for ultracold neutral atoms,'' \emph{Nature},
  vol. 462, p. 628, 2009.

\bibitem{Ray2014}
M.~W. Ray, E.~Ruokokoski, S.~Kandel, M.~M\"{o}tt\"{o}nen, and D.~S. Hall,
  ``Observation of dirac monopoles in a synthetic magnetic field,''
  \emph{Nature}, vol. 505, p. 657, 2014.

\bibitem{Lu2014}
L.~Lu, J.~Joannopoulos, and M.~Soljačić, ``Topological photonics,''
  \emph{Nat. Photons.}, vol.~8, p. 821, 2014.

\bibitem{RMP}
M.~Aspelmeyer, T.~J. Kippenberg, and F.~Marquardt, ``Cavity optomechanics,''
  \emph{Rev. Mod. Phys.}, vol.~86, pp. 1391--1452, 2014.

\bibitem{Rosenberg2009}
J.~Rosenberg, Q.~Lin, and O.~Painter, ``Static and dynamic wavelength routing
  via the gradient optical force,'' \emph{Nature Photon.}, vol.~3, pp.
  478--483, Aug. 2009.

\bibitem{Manipatruni2009}
S.~Manipatruni, J.~T. Robinson, and M.~Lipson, ``Optical nonreciprocity in
  optomechanical structures,'' \emph{Phys. Rev. Lett.}, vol. 102, p. 213903,
  May 2009.

\bibitem{Hafezi2012}
M.~Hafezi and P.~Rabl, ``Optomechanically induced non-reciprocity in microring
  resonators,'' \emph{Opt. Express}, vol.~20, no.~7, pp. 7672--7684, Mar. 2012.

\bibitem{ophwg2}
S.~J.~M. Habraken, K.~Stannigel, M.~D. Lukin, P.~Zoller, and P.~Rabl,
  ``Continuous mode cooling and phonon routers for phononic quantum networks,''
  \emph{New J. Phys.}, vol.~14, p. 115004, 2012.

\bibitem{Wang2015}
Z.~Wang, L.~Shi, Y.~Liu, X.~Xu, and X.~Zhang, ``Optical nonreciprocity in
  asymmetric optomechanical couplers,'' \emph{Scientific Reports}, vol.~5, p.
  8657, Mar. 2015.

\bibitem{manybody1}
V.~Peano, C.~Brendel, M.~Schmidt, and F.~Marquardt, ``Topological phases of
  sound and light,'' \emph{Phys. Rev. X}, vol.~5, p. 031011, 2015.

\bibitem{omcc}
K.~Fang, M.~H. Matheny, X.~Luan, and O.~Painter, ``Optical transduction and
  routing of microwave phonons in cavity-optomechanical circuits,''
  \emph{Nature Photon.}, vol.~10, p. 489–496, Jun. 2016.

\bibitem{Kim2015}
J.~Kim, M.~C. Kuzyk, K.~Han, H.~Wang, and G.~Bahl, ``Non-reciprocal brillouin
  scattering induced transparency,'' \emph{Nat. Phys.}, vol.~11, pp. 275--280,
  Jan. 2015.

\bibitem{Dong2016}
Z.~Shen, Y.-L. Zhang, Y.~Chen, C.-L. Zou, Y.-F. Xiao, X.-B. Zou, F.-W. Sun,
  G.-C. Guo, and C.-H. Dong, ``Experimental realization of optomechanically
  induced non-reciprocity,'' \emph{arXiv:1604.02297}, 2016.

\bibitem{Ruesink2016}
F.~Ruesink, M.-A. Miri, A.~Al\'{u}, and E.~Verhagen, ``Nonreciprocity and
  magnetic-free isolation based on optomechanical interactions,''
  \emph{arXiv:1607.07180}, Jul. 2016.

\bibitem{reservoir}
A.~Metelmann and A.~Clerk, ``Nonreciprocal photon transmission and
  amplification via reservoir engineering,'' \emph{Phys. Rev. X}, vol.~5, p.
  021025, 2015.

\bibitem{Abdo2013}
B.~Abdo, K.~Sliwa, L.~Frunzio, and M.~Devoret, ``Directional amplification with
  a josephson circuit,'' \emph{Phys. Rev. X}, vol.~3, p. 031001, Jul. 2013.

\bibitem{Abdo2014}
B.~Abdo, K.~Sliwa, S.~Shankar, M.~Hatridge, L.~Frunzio, R.~Schoelkopf, and
  M.~Devoret, ``Josephson directional amplifier for quantum measurement of
  superconducting circuits,'' \emph{Phys. Rev. Lett.}, vol. 112, p. 167701,
  Apr. 2014.

\bibitem{Peano2016}
V.~Peano, M.~Houde, C.~Brendel, F.~Marquardt, and A.~Clerk, ``Topological phase
  transitions and chiral inelastic transport induced by the squeezing of
  light,'' \emph{Nat. Commun.}, vol.~7, p. 10779, 2016.

\bibitem{Peano2016B}
V.~Peano, M.~Houde, F.~Marquardt, and A.~A. Clerk, ``Topological quantum
  fluctuations and travelling wave amplifiers,'' \emph{arXiv:1604.04179}, Apr.
  2016.

\bibitem{ab}
K.~Fang, Z.~Yu, and S.~Fan, ``Photonic aharonov-bohm effect based on dynamic
  modulation,'' \emph{Phys. Rev. Lett.}, vol. 108, p. 153901, 2012.

\bibitem{Groeblacher2013}
S.~Gr\"{o}blacher, J.~T. Hill, A.~H. Safavi-Naeini, J.~Chan, and O.~Painter,
  ``Highly efficient coupling from an optical fiber to a nanoscale silicon
  optomechanical cavity,'' \emph{App. Phys. Lett.}, vol. 108, p. 181104, 2013.

\bibitem{Aharanov1959}
Y.~Aharanov and D.~Bohm, ``Signigicance of electromagnetic potentials in the
  quantum theory,'' \emph{Phys. Rev.}, vol. 115, pp. 485--491, 1959.

\bibitem{Deak2012}
L.~De\'{a}k and T.~F\"{u}l\"{o}p, ``Reciprocity in quantum, electromagnetic and
  other wave scattering,'' \emph{Ann. Phys.}, vol. 327, pp. 1050--1077, 2012.

\bibitem{omc}
M.~Eichenfield, J.~Chan, R.~M. Camacho, K.~J. Vahala, and O.~Painter,
  ``Optomechanical crystals,'' \emph{Nature}, vol. 462, pp. 78--82, 2009.

\bibitem{phoxonic2}
S.~Sadat-Saleh, S.~Benchabane, F.~I. Baida, M.-P. Bernal, and V.~Laude,
  ``Tailoring simultaneous photonic and phononic band gaps,'' \emph{J. Appl.
  Phys.}, vol. 106, p. 074912, 2009.

\bibitem{phoxonic3}
J.~Gomis-Bresco, D.~Navarro-Urrios, M.~Oudich, S.~El-Jallal, A.~Griol,
  D.~Puerto, E.~Chavez, Y.~Pennec, B.~Djafari-Rouhani, F.~Alzina,
  A.~Mart\'{\i}nez, and C.~S. Torres, ``A one-dimensional optomechanical
  crystal with a complete phononic band gap,'' \emph{Nat. Commun.}, vol.~5, p.
  4452, 2014.

\bibitem{shielding}
A.~H. Safavi-Naeini and O.~Painter, ``Design of optomechanical cavities and
  waveguides on a simultaneous bandgap phononic-photonic crystal slab,''
  \emph{Opt. Express}, vol.~18, pp. 14\,926--14\,943, 2010.

\bibitem{noda}
Y.~Sato, Y.~Tanaka, J.~Upham, Y.~Takahashi, T.~Asano, and S.~Noda, ``Strong
  coupling between distant photonic nanocavities and its dynamic control,''
  \emph{Nature Photon.}, vol.~6, pp. 56--61, 2012.

\bibitem{afm}
K.~Fang, X.~Luan, M.~H. Matheny, M.~L. Roukes, and O.~Painter, \emph{in
  preparation}.

\bibitem{dynamicalgauge}
S.~Walter and F.~Marquardt, ``Dynamical gauge fields in optomechanics,''
  \emph{arXiv:1510.06754}, Oct. 2015.

\bibitem{Clerk2010}
A.~Clerk, M.~Devoret, S.~Girvin, F., Marquardt, and R.~Schoelkopf,
  ``Introduction to quantum noise, measurement, and amplification,'' \emph{Rev.
  Mod. Phys.}, vol.~82, p. 1155, 2010.

\end{thebibliography}
\end{document}